\documentstyle [11pt]{article}

\begin{document}

\begin{center}
{\Large\bf
General realization of $N=4$ supersymmetric quantum 
mechanics and its applications} \\
\end{center}

\vspace{4mm}

\begin{center}
{\large Dong Ruan}\footnote{Email: dongruan@tsinghua.edu.cn}
\end{center}
{\it	Department of Physics, Tsinghua University,
Beijing 100084, P. R. China,
	Center of Theoretical Nuclear Physics,
National Laboratory of Heavy Ion Accelerator,
Lanzhou, 730000, P. R. China
	The Key Laboratory of Quantum Information and
Measurements of Ministry of Education, Tsinghua University,
Beijing 100084, P. R. China, and
}

\begin{center}
{\large Weicheng Huang}
\end{center}
{\it Institute of Applied Chemistry, Xingjiang University,
Urumqi 830046, P. R. China}

\vfill

\hfill     {Typeset by La\TeX}

\newpage

\begin{abstract}
Based upon the general supercharges which involve not only
generators $C_j$ of the Clifford algebra C(4,0) with
positive metric, but also operators of third order, $C_j C_k C_l$, 
the general form of $N=4$ supersymmetric quantum mechanics 
(SSQM), which brings out the richer structures, is realized. 
Then from them, an one-dimensional physical
realization and a new multi-dimensional physical realization
of $N=4$ SSQM are respectively obtained by solving
the constraint conditions. 
As applications, $N=4$ dynamical superconformal symmetries, 
which possess both the $N=4$ supersymmetries and the usual 
dynamical conformal symmetries, are studied in detail by 
considering two simple superpotentials $k/x$ and $\omega x$, and 
their corresponding superalgebraic structures, which are spanned
by eight fermionic generators and six bosonic generators, are 
established as well.
\end{abstract}
\vspace{5mm}



\newpage

\section{INTRODUCTION}
\label{1}
Since the idea of supersymmetry was applied to quantum
mechanical systems \cite{nicolai,witten} to discover dynamical
supersymmetry in ordinary quantum mechanics in order to
explain the degeneracies of energy spectra
extensively studies have been undertaken over the
last twenty years in many aspects such as atomic physics,
\cite{kn,zhang,rau} nuclear physics, \cite{abcd} many-body
systems, \cite{shastry,ss} and so on. According to Witten,
\cite{witten} a supersymmetric quantum mechanical system is
characterized by the existence of $N$ Hermitian supercharges
$Q^{\alpha}$ which, together with the supersymmetric
Hamiltonian $H$ of this system, satisfy the following
superalgebraic structure
\begin{eqnarray}
\begin{array}{ll}
  \{Q^{\alpha},\: Q^{\beta}\}= 2 \delta_{\alpha\,\beta} H,  
	&   \alpha,\, \beta = 1,\,2,\, ... \, N,   \cr
  (Q^{\alpha})^{\dagger} = Q^{\alpha},    
	&  [H ,\: Q^{\alpha}]=0,  
\end{array}
\label{sqma-1}
\end{eqnarray}
where $\{\hspace{2mm}, \hspace{2mm}\}$ and $[\hspace{2mm}, 
\hspace{2mm}]$ denote an anticommutator and a commutator 
respectively. We call Eq. (\ref{sqma-1}) a supersymmetric 
quantum mechanical algebra, denoted by SS($N$). 
When $N=2$, the simplest non-trivial realization of SS(2) 
was first given by Witten \cite{witten}
\begin{eqnarray}
\begin{array}{l}
    Q^1 = {1 \over 2}\{ p \sigma_1 + U(x) \sigma_2 \}, \cr
    Q^2 = {1 \over 2}\{ p \sigma_2 - U(x) \sigma_1 \}, \cr
    H^{\mbox{w}} = {1 \over 2}\{ p^2 + [U(x)]^2 + U'(x) \sigma_3 \},
\end{array}
\label{n2}
\end{eqnarray}
where $p = -i {d \over dx}$, $U'(x) \equiv {d \over dx} U(x)$,
$U(x)$ is generally called a superpotential, and
$\sigma_i$ ($i$ $=$ $1$, $2$, $3$) are the usual Pauli matrices.
\cite{messiah} This supersymmetric Hamiltonian $H^{\mbox{w}}$ describes
a quantum mechanical system of a spin-${1 \over 2}$ particle moving
on a line ($x$-axis).

While $N=2$ supersymmetric quantum mechanics (SSQM) has drawn much
attention \cite{cks,junker} due to its simpler mathematical 
structure, however, there were only a few attempts at studying 
$N=4$ SSQM, \cite{cr,ry,jll,mathur,ft} which possesses the higher 
degeneracies than the $N=2$ SSQM. 
For example, in one dimension, four-fold degeneracies of energy 
spectrum may typically occur in the $N=4$ SSQM, whereas double 
degeneracies in the $N=2$ SSQM. 
All the supercharges considered in Refs. \cite{cr,ry,jll,mathur,ft} 
have the following form
\begin{equation}
	Q^{\alpha} = {1 \over {\sqrt 2}} \sum_{j=1}^{r}
        A_{j}^{\alpha} C_j,
\hspace{4mm}
	\alpha = 1, \, 2, ..., N=4,
\label{n4-cr}
\end{equation}
where $A_{j}^{\alpha}$ are first-order differential operators
with respect to, `bosonic' degrees of freedoms, the Cartesian 
coordinates \{$x_n|n=1$, 2,..., $d$\} and the corresponding 
momentum operators 
\{$p_n = -i {\partial \over \partial x_n} \equiv -i \partial_n
|n=1$, 2,..., $d$\} 
in $d$-dimensional space,
and $C_j$, `fermionic' degrees of freedoms, are generators 
of the Clifford algebra C$(r,0)$ with positive metric 
in $r$-dimensional flat carrier space. \cite{nepomechie,okubo}
They satisfy
\begin{eqnarray}
\begin{array}{l}
  [x_n,\: p_m] = i \delta_{nm},       \hspace{4mm}
         n, \, m = 1,\, 2, ...,d;           \cr
   \{C_j, \: C_l \} = 2 \delta_{jl},  \hspace{3mm}
   C^{\dagger}_{j} = C_{j},           \hspace{3mm}
         j, \, l=1,\,2,\,...,r;       \cr
  [x_n,\: C_j] = [p_n,\: C_j] =0.
\end{array}
\label{degree}
\end{eqnarray}
Obviously, for arbitrary $N$, the supercharges (\ref{n4-cr}), 
being linear combinations of the fermionic operators $C_j$
multiplied by the bosonic operators $A_j^{\alpha}$, 
are natural generalizations of the supercharges in 
Witten's realization (\ref{n2}) of $N=2$ SSQM.

In fact, when $r \geq 4$, the Clifford algebra C$(r,0)$, 
after a graded structure introduced, may yield a superalgebra.
\cite{kac,sh} The generators $C_j$ of C$(r,0)$, together with 
operators of odd orders in $C_j$, span the `odd' space
of this superalgebra, in which anticommuting operations 
among all these `odd' elements are allowed.
For example, the odd space of the superalgebra associated with 
C(4,0) is spanned by the odd elements $C_j$ ($j=1$, 2, 3, 4)
and $C_j C_k C_l$ ($1 \leq j < k < l \leq 4$).
The purpose of this paper is to realize the general form
of $N=4$ SSQM in arbitrary dimension starting from the general 
supercharges in which the fermionic degrees of freedoms include
all the odd elements of the superalgebra associated with C(4,0). 
As we shall see below, this realization brings out the richer structures. 

This paper is arranged as follows. In Sec. \ref{2}, the general
form of $N=4$ SSQM is studied in detail by means of the Clifford
algebras C(4,0) and C(0,3). 
In Sec. \ref{3}, an one-dimensional physical realization and 
a new multi-dimensional physical realization for the $N=4$ SSQM
are respectively obtained by solving the constraint conditions. 
In Sec. \ref{4}, as applications, $N=4$ superconformal quantum 
mechanics in one dimension, which is expanded from 
the one-dimensional realization of $N=4$ SSQM obtained in 
Sec. \ref{3}, is discussed in detail by considering two simple 
superpotentials $k/x$ and $\omega x$, and their corresponding 
superalgebraic structures are established.
A simple summary is given in the final section. 

Throughout this paper we shall adopt units wherein 
$\hbar = m = 1$, the symbol $[x]$ means taking an
integer part of the real number $x$, and T in the expression 
$A^{\mbox{T}}$ is referred to as transpose of the matrix $A$.

\section{GENERAL FORM OF $N=4$ SSQM}
\label{2}
For $N=4$, the four supercharges take the following general form
\begin{eqnarray}
\begin{array}{l}
Q^{\alpha} = {1 \over {\sqrt 2}} \left( \sum\limits_{j}
         A_j^{\alpha} C_j +  {i \over 3!} \sum\limits_{jklm}
          \epsilon_{jklm}  D_j^{\alpha} C_k C_l C_m 
	\right),  \cr
     \alpha, \, j,\, k, \, l, \, m = 1, \, 2, \,3, \,4,
\end{array}
\label{n4-q}
\end{eqnarray}
where $\epsilon_{jklm}$ is a four-dimensional Levi-Civita symbol, 
$C_j$ are the generators of the Clifford algebra C(4,0), 
$A_j^{\alpha}$ and $D_j^{\alpha}$ are the Hermitian first-order
differential operators of the $d$-dimensional coordinates 
\{$x_n$\} and momentum operators \{$p_n$\}. 
Clearly, the supercharges (\ref{n4-cr}) are the special cases 
of those given by Eq. (\ref{n4-q}) with setting $D_j^{\alpha}=0$. 

Substituting Eq. (\ref{n4-q}) into the first equation of 
Eq. (\ref{sqma-1}), we may obtain the corresponding supersymmetric 
Hamiltonian
\begin{equation}
  H= {1 \over 2}(U + V C_1 C_2 C_3 C_4) +
     {1 \over 2} \sum_{l}^{q}\, \sum_{j\,k}\,
                 \wp^{l}_{jk} B_l \Gamma_{jk},
\label{h-1}
\end{equation}
where
\begin{eqnarray}
\begin{array}{ll}
1. &  U= \sum\limits_j
               [(A_{j}^{\alpha})^2 + (D_{j}^{\alpha})^2],
        \hspace{4mm}
          \mbox{for any}\;   \alpha;             \cr
2. &  V= i \sum\limits_j
               [A_{j}^{\alpha}, \: D_{j}^{\alpha}],
        \hspace{4mm}
          \mbox{for any}\; \alpha;               \cr
3. &  i [A_{j}^{\alpha}, \: A_{k}^{\alpha}] +
        i [D_{j}^{\alpha}, \: D_{k}^{\alpha}] +
        {1 \over 2} \sum\limits_{mn}\, \epsilon_{jkmn}
        ( \{A_{m}^{\alpha}, \:D_{n}^{\alpha} \}    \cr
{} &     \hspace{3mm}
       - \{A_{n}^{\alpha}, \:D_{m}^{\alpha} \} )
       = - \sum\limits_{l=1}^q\, \wp_{jk}^{l} B_l,
        \hspace{4mm} \mbox{for any} \;\alpha;    \cr
4. &   \sum\limits_j\,
        (\{A_{j}^{\alpha}, \: A_{j}^{\beta} \}
         +  \{D_{j}^{\alpha}, \: D_{j}^{\beta} \} ) =0,
        \hspace{4mm} \alpha \not= \beta;     \cr
5. &   [A_{j}^{\alpha}, \: A_{k}^{\beta}]
          -[A_{k}^{\alpha}, \: A_{j}^{\beta}]
      + [D_{j}^{\alpha}, \: D_{k}^{\beta}]
          -[D_{k}^{\alpha}, \: D_{j}^{\beta}]  \cr
{} &  \hspace{3mm}
       =- i \sum\limits_{mn}\, \epsilon_{jkmn}
         ( \{A_{n}^{\alpha}, \:D_{m}^{\beta} \}
           + \{A_{n}^{\beta}, \:D_{m}^{\alpha} \} ),
        \hspace{4mm} \alpha \not= \beta;        \cr
6. & \sum\limits_j
         ([ A_{j}^{\alpha}, \:D_{j}^{\beta} ]
           + [A_{j}^{\beta}, \:D_{j}^{\alpha} ] )=0,
        \hspace{4mm} \alpha \not= \beta;        \cr
7. &  \Gamma_{jk} \equiv {i \over 4} [C_j, \: C_k].
\end{array}
\label{condition}
\end{eqnarray}
In Eq. (\ref{h-1}), $q$ antisymmetric matrices
$\wp^l$ ($\wp^l_{jk}$ $=$ $-\wp^l_{kj}$) and Hermitian
operators $B_l$ have to be determined by the third equation
in Eq. (\ref{condition}).

Now introduce the following linear transformations
\begin{eqnarray}
\begin{array}{lll}
     A_{i}^{\bar{\alpha}}= \Xi_{ij}^{\bar{\alpha}}A_{j}^{4},
	&  A_{j}^{4} \equiv A_{j},  
	&  \bar{\alpha}= 1,\,2,\,3;     \cr
     D_{i}^{\bar{\alpha}}= \Xi_{ij}^{\bar{\alpha}}D_{j}^{4},
	&  D_{j}^{4} \equiv D_{j}, 
	&  {}
\end{array}
\label{a-d}
\end{eqnarray}
In order to satisfy the first, second, fourth and sixth equations
in Eq. (\ref{condition}), $\Xi^{\bar{\alpha}}$ should be real 
antisymmetric, 
\begin{equation}
   (\Xi^{\bar{\alpha}})^{\mbox{T}} = (\Xi^{\bar{\alpha}})^{-1} =
    - \Xi^{\bar{\alpha}},
\label{o1}
\end{equation}
and satisfy
\begin{equation}
   \{ \Xi^{\bar{\alpha}},  \;  \Xi^{\bar{\beta}} \} =
    - 2 \delta_{\bar{\alpha}\, \bar{\beta}},
\label{o2}
\end{equation}
that is, $\Xi^{\bar{\alpha}}$ constitute the Clifford algebra C(0,3)
with negative metric.
The third and fifth equations in Eq. (\ref{condition}) require 
further that $\Xi^{\bar{\alpha}}$ and $\wp^l$ satisfy 
\begin{eqnarray}
{} & \Xi_{jk}^{\bar{\alpha}} \epsilon_{jkmn} =
            -2 \Xi_{mn}^{\bar{\alpha}},       \\
{} & [\Xi^{\bar{\alpha}}, \: \wp^l]=0.
\label{t-e}
\end{eqnarray}
Eq. (\ref{t-e}) shows that the number of antisymmetric 
matrices $\wp^l$ is three, i.e., $l=1$, 2, 3.
It follows that matrix representations of 
$\Xi^{\bar{\alpha}}$ and $\wp^l$ that satisfy Eqs. (\ref{o1}), 
(\ref{o2}) and (\ref{t-e}) may be taken as \cite{cr,okubo}
\begin{eqnarray}
  \Xi^1= \left( \begin{array}{cc} i \sigma_2  &  0 \cr
                                0  &  i \sigma_2 \cr
              \end{array}
       \right),
  \hspace{4mm}
  \Xi^2= \left( \begin{array}{cc} 0  &  \sigma_3 \cr
                                - \sigma_3  &  0\cr
              \end{array}
       \right),
  \hspace{4mm}
  \Xi^3= \left( \begin{array}{cc} 0  &  \sigma_1 \cr
                                -\sigma_1  &  0 \cr
              \end{array}
       \right);
\label{o-m}
\end{eqnarray}
\begin{eqnarray}
  \wp^1= \left( \begin{array}{cc} 0  &  i \sigma_2  \cr
                                 i \sigma_2  &   0   \cr
                 \end{array}
       \right),
  \hspace{4mm}
  \wp^2= \left( \begin{array}{cc} 0  &  I \cr
                                 - I  &  0  \cr
                 \end{array}
       \right),
  \hspace{4mm}
  \wp^3= \left( \begin{array}{cc} -i\sigma_2 & 0       \cr
                                      0  &   i \sigma_2 \cr
                 \end{array}
       \right).
\label{f-m}
\end{eqnarray}

With the help of Eqs. (\ref{a-d}) and (\ref{o-m}), 
the four supercharges (\ref{n4-q}), in which the differential
operators $A_j$ and $D_j$ now are independent of the index 
$\alpha$, read
\begin{eqnarray}
\begin{array}{rl}
Q^1 = &  {1 \over {\sqrt 2}}
             [ A_2 C_1 - A_1 C_2 + A_4 C_3 - A_3 C_4      \cr
 {}   &  + i (D_2 C_2 C_3 C_4 + D_1 C_3 C_4 C_1
              + D_4 C_1 C_2 C_4 + D_3 C_1 C_2 C_3 )],     \cr
Q^2 = &  {1 \over {\sqrt 2}}
             [ A_3 C_1 - A_4 C_2 - A_1 C_3 + A_2 C_4      \cr
 {}   &  + i (D_3 C_2 C_3 C_4 + D_4 C_3 C_4 C_1
              - D_1 C_1 C_2 C_4 - D_2 C_1 C_2 C_3 )],     \cr
Q^3 = &  {1 \over {\sqrt 2}}
             [ A_4 C_1 + A_ 3C_2 - A_2 C_3 - A_1 C_4      \cr
 {}   &  + i (D_4 C_2 C_3 C_4 - D_3 C_3 C_4 C_1
             - D_2 C_1 C_2 C_4 + D_1 C_1 C_2 C_3 )],      \cr
Q^4 = &  {1 \over {\sqrt 2}}
             [ A_1 C_1 + A_2 C_2 + A_3 C_3 + A_4 C_4      \cr
 {}   &  + i (D_1 C_2 C_3 C_4 - D_2 C_3 C_4 C_1
              + D_3 C_1 C_2 C_4 - D_4 C_1 C_2 C_3 )].
\label{q-exp1}
\end{array}
\end{eqnarray}
By making use of Eqs. (\ref{a-d}), (\ref{o-m}) and (\ref{f-m}),
we can obtain from the third equation in Eq. (\ref{condition})
constraint conditions that $A_j$ and $D_j$ need satisfying
\begin{eqnarray}
\begin{array}{ll}
B_1 & = -i[A_1,\, A_4] - i[D_1,\, D_4] - \{ A_2,\, D_3\}
        + \{ A_3,\, D_2\}                               \\
{}  & = +i[A_2,\, A_3] + i[D_2,\, D_3] + \{ A_1,\, D_4\}
        - \{ A_4,\, D_1\},                              \\
B_2 & = -i[A_2,\, A_4] - i[D_2,\, D_4] + \{ A_1,\, D_3\}
        - \{ A_3,\, D_1\}                               \\
{}  & = -i[A_1,\, A_3] - i[D_1,\, D_3] + \{ A_2,\, D_4\}
        - \{ A_4,\, D_2\},                              \\
B_3 & = -i[A_3,\, A_4] - i[D_3,\, D_4] - \{ A_1,\, D_2\}
        + \{ A_2,\, D_1\}                               \\
{}  & = +i[A_1,\, A_2] + i[D_1,\, D_2] + \{ A_3,\, D_4\}
        - \{ A_4,\, D_3\}.
\label{cc}
\end{array}
\end{eqnarray}
Correspondingly, the supersymmetric Hamiltonian (\ref{h-1})
becomes
\begin{eqnarray}
\begin{array}{ll}
H =&  {1 \over 2} \sum\limits_{j=1}^{4} \, [A^2_j + D^2_j]
      + [ B_1 (\Gamma_{14} + \Gamma_{32}) +
          B_2 (\Gamma_{24} + \Gamma_{13})    \cr
{} &    + B_3 (\Gamma_{34} + \Gamma_{21}) ]
      + {i\over 2} \sum\limits_{j=1}^{4}\, [A_j,\, D_j]
	 C_1 C_2 C_3 C_4.
\label{h-2}
\end{array}
\end{eqnarray}

Similar to $N=2$ SSQM, \cite{cks} we may further rewrite 
the four supercharges (\ref{q-exp1}) in the raising/lowering 
form, which are closely related to the factorization method, 
\cite{ih}
\begin{eqnarray}
\begin{array}{l}
      Q^{\pm}_1 = {1 \over \sqrt{2}} (Q_4 \mp i Q_1),
      \hspace{4mm}
      Q^{\pm}_2 = {1 \over \sqrt{2}} (Q_2 \mp i Q_3),
\end{array}
\label{q2}
\end{eqnarray}
and satisfy $Q^{\mp}_{\mu} = (Q^{\pm}_{\mu})^{\dagger}$ 
($\mu=1$, 2), after redefining the four generators $C_j$ 
($j=1$, 2, 3, 4) of C(4,0) as 
\begin{eqnarray}
      C^{\pm}_1 = {1 \over \sqrt{2}} (C_1 \pm i C_2),
      \hspace{4mm}
      C^{\pm}_2 = {1 \over \sqrt{2}} (C_3 \pm i C_4).
\label{c2}
\end{eqnarray}
Thus, the four supercharges (\ref{q-exp1}) and the supersymmetric 
Hamiltonian (\ref{h-2}) can be written respectively in the forms
\begin{eqnarray}
\begin{array}{ll}
 Q^{+}_1 = & (A_1 - i A_2)C^{+}_{1} + (A_3 - i A_4) C^{+}_{2}
             -i (D_1 - i D_2)
                [C^{+}_{2},\, C^{-}_{2}] C^{+}_{1}     \\
       {}  & +i (D_3 - i D_4)
                [C^{+}_{1},\, C^{-}_{1}] C^{+}_{2}\,, \\
 Q^{-}_1 = & (A_1 + i A_2)C^{-}_{1} + (A_3 + i A_4) C^{-}_{2}
             +i (D_1 + i D_2)
                [C^{+}_{2},\, C^{-}_{2}] C^{-}_{1}   \\
       {}  & -i (D_3 + i D_4)
                [C^{+}_{1},\, C^{-}_{1}] C^{-}_{2}\,, \\
 Q^{+}_2 = & (A_3 - i A_4)C^{-}_{1} - (A_1 - i A_2) C^{-}_{2}
             -i (D_3 - i D_4)
                [C^{+}_{2},\, C^{-}_{2}] C^{-}_{1}     \\
       {}  & + i (D_1 - i D_2)
                [C^{+}_{1},\, C^{-}_{1}] C^{-}_{2}\,, \\
 Q^{-}_2 = & (A_3 + i A_4)C^{+}_{1} - (A_1 + i A_2) C^{+}_{2}
               +i (D_3 + i D_4)
                [C^{+}_{2},\, C^{-}_{2}] C^{+}_{1}   \\
       {}  & - i (D_1 + i D_2)
                [C^{+}_{1},\, C^{-}_{1}] C^{+}_{2},
\label{sc}
\end{array}
\end{eqnarray}
and 
\begin{eqnarray}
\begin{array}{ll}
H =&  {1 \over 2} \sum\limits_{j=1}^{4} \, [A^2_j + D^2_j]
    + \left\{ - B_1 (C^{+}_1 C^{-}_2 - C^{-}_1 C^{+}_2)
        +  i B_2 (C^{+}_1 C^{-}_2 + C^{-}_1 C^{+}_2) \right.   \\
{} &   + \left. {1 \over 2} B_3 ([C^{+}_1,\, C^{-}_1 ]
           - [C^{+}_2,\, C^{-}_2]) \right\}
    - {i\over 2} \sum\limits_{j=1}^{4}\, [A_j,\, D_j]
	[C^{+}_1,\, C^{-}_1] [C^{+}_2,\, C^{-}_2].
\label{h-3}
\end{array}
\end{eqnarray}
It is easy to check that Eqs. (\ref{sc}) and (\ref{h-3}) satisfy
SS(4) [see Eq. (\ref{sqma-1})], which now becomes
\begin{eqnarray}
\begin{array}{ll}
   \{ Q^{+}_{\mu},\: Q^{-}_{\nu} \}=
       2 \delta_{\mu\nu} H,
&  \mu,\nu = 1,\,2 , \cr
   \{ Q^{\pm}_{\mu},\: Q^{\pm}_{\nu} \} =0,
&  [H ,\: Q^{\pm}_{\mu}]=0.
\label{sqma-2}
\end{array}
\end{eqnarray}

\section{PHYSICAL REALIZATIONS OF $N=4$ SSQM}
\label{3}
By virtue of the results obtained in Sec. \ref{2}, we shall discuss
in this section two physical realizations of $N=4$ SSQM through
choosing the concrete forms for $A_j$ and $D_j$ in Eqs. (\ref{sc}) 
and (\ref{h-3}) to ensure the kinetic energy and potential energy 
terms appear in the corresponding $N=4$ supersymmetric Hamiltonian. 

\subsection{ONE-DIMENSIONAL REALIZATION}
\label{3-1}
Take
\begin{equation}
  A_4 = p = -i {d \over dx};  \hspace{4mm}
  D_{\bar{n}} = 0, \hspace{4mm} \bar{n}=1,\,2,\,3,
\label{ad-1d}
\end{equation}
and the other components $A_{\bar{n}}$ and $D_4$ are real functions
of $x$. Substituting them into the constraint conditions (\ref{cc}) 
gives rise to
\begin{equation}
   D_4 = {A^{\prime}_1 \over 2 A_1} =
         {A^{\prime}_2 \over 2 A_2} =
         {A^{\prime}_3 \over 2 A_3},
\label{da-cond}
\end{equation}
where the symbol ``$\prime$" means derivation with respect to $x$.
In order to satisfy Eq. (\ref{da-cond}), we may choose for simplicity
\begin{equation}
   A_{\bar{n}} = k_{\bar{n}} W,
\label{an}
\end{equation}
where $k_{\bar{n}}$ are constants, and $W$, here referred to 
as a superpotential, is an `arbitrary' real function of $x$.
Accordingly, we have
\begin{equation}
   D_4 = {W' \over 2 W}.
\label{d4}
\end{equation}
It follows by inserting Eqs. (\ref{an}) and (\ref{d4}) into 
Eqs. (\ref{sc}) and (\ref{h-3}) that the four supercharges 
and $N=4$ supersymmetric Hamiltonian in one dimension 
have respectively the following forms
\begin{eqnarray}
\begin{array}{l}
 Q^{+}_1 = (-i p + k_3 W )C^{+}_{2} + k^{-} W C^{+}_{1}
     +{W' \over 2W}[C^{+}_{1},\, C^{-}_{1}] C^{+}_{2},  \\
 Q^{-}_1 = (+i p + k_3 W )C^{-}_{2} + k^{+} W C^{-}_{1}
     +{W' \over 2W}[C^{+}_{1},\, C^{-}_{1}] C^{-}_{2},  \\
 Q^{+}_2 = (-i p + k_3 W )C^{-}_{1} - k^{-} W C^{-}_{2}
     -{W' \over 2W}[C^{+}_{2},\, C^{-}_{2}] C^{-}_{1},  \\
 Q^{-}_2 = (+i p + k_3 W )C^{+}_{1} - k^{+} W C^{+}_{2}
     -{W' \over 2W}[C^{+}_{2},\, C^{-}_{2}] C^{+}_{1},
\label{q-1d}
\end{array}
\end{eqnarray}
where $k^{\pm} \equiv k_1 \pm i k_2$, and
\begin{eqnarray}
\begin{array}{rl}
H =& {1 \over 2} p^2 + {1 \over 2} k^2 W^2 +
     {1 \over 2} \left( {W' \over 2 W} \right)^2
     + \left\{  - k_1 W' (C^{+}_1 C^{-}_2 -
                     C^{-}_1 C^{+}_2) \right.            \\
{} &  \left. +  i k_2  W' (C^{+}_1 C^{-}_2 +
                    C^{-}_1 C^{+}_2)
      + {1 \over 2} k_3  W'
                   ([C^{+}_1,\, C^{-}_1 ] -
                    [C^{+}_2,\, C^{-}_2]) \right\}       \\
{} &  - {1 \over 2} [C^{+}_1,\, C^{-}_1]
                    [C^{+}_2,\, C^{-}_2] \,
        \left( {W' \over 2 W} \right)' ,                 \\
\equiv &  {1 \over 2} p^2 + V(x;C_j),
\end{array}
\label{h-1d}
\end{eqnarray}
with $k^2= k_1^2 + k_2^2 + k_3^2$. Obviously, the Hamiltonian
(\ref{h-1d}) possesses a usual kinetic energy term ${1 \over 2} p^2$
and a potential function $V(x;C_j)$, so this resulting realization,
Eqs. (\ref{q-1d}) and (\ref{h-1d}), may be applied to the real
quantum mechanical systems provided that $W$ and $C_j$ are 
appropriately taken.

For the sake of convenient applications, let us further discuss 
the explicit matrix form for the one-dimensional $N=4$ SSQM given by 
Eqs. (\ref{q-1d}) and (\ref{h-1d}). 
In fact the Clifford algebra C(4,0) is isomorphic to the well-known 
Dirac algebra in the relativistic quantum mechanics, \cite{bd} 
here we may take the following matrix representation for $C_j$
\begin{eqnarray}
      C_{\bar{m}}
          =  \left( \begin{array}{cc}
                  0                 &  i \sigma_{\bar{m}} \cr
                  -i \sigma_{\bar{m}}   &  0
                     \end{array}
              \right ),
   \hspace{4mm}
       \bar{m}=1,\,2,\,3,
   \hspace{4mm}
      C_4 = \left( \begin{array}{cc}
                  0                &    I_{2 \times 2} \cr
                  I_{2 \times 2}   &    0
                     \end{array}
              \right ),
\label{c4}
\end{eqnarray}
where $I_{2 \times 2}$ is a $2 \times 2$ unit matrix, 
then the four supercharges (\ref{q-1d})
and the supersymmetric Hamiltonian (\ref{h-1d})
read respectively
\begin{eqnarray}
 Q^{+}_{1} = \left( \begin{array}{cccc}
        0 & 0                & \eta^{+}   &  \varepsilon^{-}  \cr
        0 & 0                &        0   &  0            \cr
        0 & -\varepsilon^{-} &        0   &  0            \cr
        0 & \zeta^{+}        &        0   &  0
                    \end{array}
            \right),
\hspace{4mm}
Q^{+}_{2} = \left( \begin{array}{cccc}
        0               &  0 &       0  &  0               \cr
        0               &  0 & \eta^{+} &  \varepsilon^{-} \cr
        \varepsilon^{-} &  0 &       0  &  0               \cr
        -\zeta^{+}      &  0 &       0  &  0
                    \end{array}
            \right),
\label{q-1d-matrix}
\end{eqnarray}
where $ \eta^{\pm} \equiv p \pm i (k_3 W + { W' \over 2 W })$,
$\zeta^{\pm} \equiv p \pm i (k_3 W - { W' \over 2 W })$,
$\varepsilon^{\pm} \equiv \mp i k^{\pm} W$, and  
\begin{eqnarray}
\begin{array}{rl}
    H  =  &  {1 \over 2} p^2 + {1 \over 2} k^2 W^2
            + {1 \over 2}
                \left( {W^{\prime} \over 2 W} \right)^2   \cr
      {}  & + {1 \over 2}
  \left (
  \begin{array}{cccc}
      -\left( {W' \over 2 W} \right)'     & {}  & {}  & {}  \\
      {} & -\left( {W' \over 2 W} \right)'      & {}  & {}  \\
      {} & {} & \left( {W' \over 2 W} \right)' + 2 k_3 W'
              & 2 k^{-} W'                                  \\
      {} & {} & 2 k^{+} W'
              & \left( {W' \over 2 W} \right)' - 2 k_3 W'
  \end{array}
 \right ),
\end{array}
\label{h-1d-matrix}
\end{eqnarray}
A comparison between Eq. (\ref{h-1d-matrix}) and the third equation
in Eq. (\ref{n2}) shows that in one dimension the matrix form 
of $N=4$ supersymmetric Hamiltonian is quasi-diagonal, 
whereas the one of $N=2$ supersymmetric Hamiltonian is completely 
diagonal [The quasi-diagonal form of $N=2$ supersymmetric Hamiltonian  
appear in multi-dimension only, for example, see Ref. \cite{abei}.] 

Especially, when $k^{\pm}=0$, it follows from Eq. (\ref{h-1d-matrix}) 
that we may obtain a completely diagonal $N=4$ supersymmetric Hamiltonian
\begin{eqnarray}
\begin{array}{rl}
    \ddot{H} = & {1 \over 2} p^2 + {1 \over 2} k_3^2 W^2
         + {1 \over 2} \left( {W^{\prime} \over 2 W} \right)^2   \cr
      {}  & + {1 \over 2}
  \left (
  \begin{array}{cccc}
      -\left( {W' \over 2 W} \right)'     & {}  & {}  & {}  \\
      {} & -\left( {W' \over 2 W} \right)'      & {}  & {}  \\
      {} & {} & \left( {W' \over 2 W} \right)' + 2 k_3 W'
              & {}                                  \\
      {} & {} & {}
              & \left( {W' \over 2 W} \right)' - 2 k_3 W'
  \end{array}
  \right )             \\
    \equiv &  \mbox{diag}[\ddot{H}_1, \; \ddot{H}_2, \;
    \ddot{H}_3, \; \ddot{H}_4].
\end{array}
\label{diagonal} 
\end{eqnarray}
Note that in Eq. (\ref{diagonal}) the first and second diagonal component
Hamiltonians $\ddot{H}_1$ and $\ddot{H}_2$ are identical, but neither of
them can be abandoned because of requirement of SS(4). 
The corresponding supercharges $\ddot{Q}^{\pm}_{\mu}$ ($\mu=1$, 2) may be
directly obtained by setting $k^{\pm}=0$ in Eq. (\ref{q-1d-matrix}).
Since $\ddot{H}$ commutes with $\ddot{Q}^{\pm}_{\mu}$,
the energy spectra of $\ddot{H}_i$ ($i=1$, 2, 3, 4) are identical 
except for the ground state, which is also an elementary property of 
SSQM. Consequently, $\ddot{H}_i$ are called superpartner Hamiltonians

Let four-component spinor eigenfunction of $\ddot{H}$ be
$\ddot{\psi}=[\ddot{\psi}_{1},\: \ddot{\psi}_{2},\: \ddot{\psi}_{3},
\: \ddot{\psi}_{4}]^{\mbox{T}}$, in which $\ddot{\psi}_{i}$ 
($i=1$, 2, 3, 4) are respectively eigenfunctions of $\ddot{H}_i$ 
belonging to energy eigenvalues $\ddot{E}_i$.  
In terms of the third equation in Eq. (\ref{sqma-2}),
the four eigenfunctions $\ddot{\psi}_{1}$, $\ddot{\psi}_{2}$, 
$\ddot{\psi}_{3}$, $\ddot{\psi}_{4}$ may be related by 
the four supercharges $\ddot{Q}^{\pm}_{\mu}$, or, more concretely, 
by the four first-order differential operators 
$\eta^{\pm}$ and $\zeta^{\pm}$ given in Eq. (\ref{q-1d-matrix})
\begin{equation}
	\ddot{\psi}_{3} \stackrel{\eta^+}{\longrightarrow} 
	\ddot{\psi}_{1} = \ddot{\psi}_{2}
	\stackrel{\zeta^+}{\longrightarrow}
	\ddot{\psi}_{4}
 \hspace{3mm} \mbox{and} \hspace{3mm}
	\ddot{\psi}_{3} \stackrel{\eta^-}{\longleftarrow} 
	\ddot{\psi}_{1} = \ddot{\psi}_{2}
	\stackrel{\zeta^-}{\longleftarrow}  
	\ddot{\psi}_{4}.
\label{trans-wfs}
\end{equation}
Similar to $N=2$ case, \cite{cks} the $N=4$ supersymmtery of
some $N=4$ supersymmetric quantum mechanical system is broken 
if this system has no zero-energy ground state, and is unbroken
if this system has a zero-energy ground state.
A typical structure of the four-fold degenerate energy spectrum 
of $\ddot{H}$ is illustratively depicted in Fig. 1, which 
$N=4$ supersymmetry is broken.

Furthermore, introduce
\begin{eqnarray}
  X_3={1 \over 2} \sum_{\mu} [C^{+}_{\mu}, \: C^{-}_{\mu}],
\hspace{4mm}
  X^{\pm} = \mp C^{\pm}_{1} C^{\pm}_{2},
\label{b-matrix}
\end{eqnarray}
which satisfy
\begin{equation}
    [X_3 , \: X^{\pm}] = \pm 2 X^{\pm},  \hspace{4mm}
    [X^{+} , \: X^{-}] =  X_3,
\end{equation}
that is, $X_3$ and $X^{\pm}$ span an internal SO(3) algebra.
Due to the fact $[ \ddot{H}, \: X_3 ] = 0$,
we may use the values of $X_3$ to label the energy spectra of
$\ddot{H}_i$ ($i=1$, 2, 3, 4). 
With the help of Eq. (\ref{c4}), the matrix representation 
of $X_3$ is
\begin{eqnarray}
   X_3 =  \left (
            \begin{array}{llll}
              1  & {} & {} & {}  \\
              {} & -1 & {} & {}  \\
              {} & {} & 0  & {}   \\
              {} & {} & {} & 0
            \end{array}
         \right ).
\label{b3}
\end{eqnarray}
Denoting the values of $X_3$ by $\beta$ $=$ $1$, $-1$, $0$
and $0$, then the energy spectrum of $\ddot{H}_{1}$ belongs to
the $\beta = 1$ sector, that of $\ddot{H}_{2}$ to
the $\beta = -1$ sector and so on. 
Hence, though the first and second superpartner Hamiltonians 
$\ddot{H}_{1}$ and $\ddot{H}_{2}$ are identical, their energy spectra 
belong to the different sectors respectively; though the third and forth
superpartner Hamiltonians $\ddot{H}_{3}$ and $\ddot{H}_{4}$ are different,
their energy spectra belong to the same $\beta = 0$ sector.

\subsection{NEW MULTI-DIMENSIONAL REALIZATION}

The $N=4$ SSQM obtained in Sec. \ref{2} itself is valid for 
the arbitrary dimensions. In this subsection we shall put forward 
a new multi-dimensional physical realization of $N=4$ SSQM by taking 
the following matrix forms for the Hermitian operators $A_4$ and 
$D_4$ in Eq. (\ref{sc}) 
\begin{eqnarray}
  A_4 = \sum_{j=1}^{d} (p_j + L_j) \tau_j,
     \hspace{4mm}
  D_4 = \sum_{j=1}^{d}  F_j \tau_j,
\label{ad-multi}
\end{eqnarray}
and the other components, $D_{\bar{n}} = 0$ ($\bar{n}= 1$, 2, 3), 
$A_{\bar{n}}$, together with $L_j$ and $F_j$ in Eq. (\ref{ad-multi}), 
are the real functions of the coordinates \{$x_n$\} in $d$-dimensional 
space. Here, $\tau_j$ are a set of Hermitian matrices which we assume
to commute with $p_j$, $L_j$, $F_j$ and $C_j$. 
The fact that $\tau_j$ commute with $C_j$
implies that they should be respectively considered as
\begin{eqnarray}
   \tau_j \sim I_{4 \times 4} \otimes  \tau_j,   \hspace{1cm}
   C_j  \sim C_j  \otimes I_{t \times t },
\end{eqnarray}
where the subscript $4$ is the dimension of the matrix representation
of C(4,0), and the subscript $t$ stands for the order of the matrices
$\tau_j$. Note that in the present realization the number of matrices 
$\tau_j$ is equal to the dimensions of space.
In order to produce the usual kinetic energy term, we may,
after substituting Eq. (\ref{ad-multi}) into Eq. (\ref{h-3}),
take $\tau_j$ ($j = 1$, 2, ..., $d$) so that
\begin{eqnarray}
  \{ \tau_j, \;   \tau_l \} = 2 \delta_{jl},
\label{tau}
\end{eqnarray}
that is, $\tau_j$ constitute the Clifford algebra C$(d,0)$ 
as well.

The constraint conditions (\ref{cc}) require that
\begin{eqnarray}
  2 A_{\bar{n}} F_j = \partial_j A_{\bar{n}},
  \hspace{4mm}    \bar{n}=1, 2, 3,   \hspace{4mm}
  j= 1, 2,...,d.
\label{solvable}
\end{eqnarray}
Similar to the one-dimensional case [see Eq. (\ref{an})],
a simple choice for Eq. (\ref{solvable}) is
\begin{eqnarray}
  A_{\bar{n}} = k_{\bar{n}} W, 
\label{a}
\end{eqnarray}
where $k_{\bar{n}}$ are constants, the superpotential $W$
is a real function of \{$x_n$\}. In consequence, we have
\begin{eqnarray}
  F_j = \frac{\partial_j W}{2 W}.
\label{k}
\end{eqnarray}

It follows by substituting Eqs. (\ref{ad-multi}), (\ref{a}) and (\ref{k})
into Eq. (\ref{h-3}) that we have the $N=4$ supersymmetric Hamiltonian in 
$d$-dimensional space
\begin{eqnarray}
\begin{array}{ll}
    H =  & {1 \over 2} \sum\limits_{j}^{d}
             (-i \partial_j + L_j)^2
          + {1 \over 2} k^2 W^2
          + {1 \over 2} \sum\limits_{j}^{d}
              \left( \frac{\partial_j W}{2 W} \right)^2
          - {1 \over 2} \sum\limits_{k<l}^{d}
              {\cal F}_{kl} \tau_{kl}                             \\
     {}  & + \sum\limits_{j}^{d}  \tau_j (\partial_j W)
           \left\{ -  k_1 ( C_{1}^{+} C_{2}^{-} -
                       C_{1}^{-} C_{2}^{+})
              + i k_2 ( C_{1}^{+} C_{2}^{-} +
                        C_{1}^{-} C_{2}^{+})   \right.         \\
     {}  & \left.  +  {1 \over 2} k_3 ( [ C_{1}^{+}, \: C_{1}^{-} ]-
                        [ C_{2}^{+}, \: C_{2}^{-} ] )
           \right\}
            - {1 \over 2} 
               [ C_{1}^{+}, \: C_{1}^{-} ]
               [ C_{2}^{+}, \: C_{2}^{-} ]
           \left\{   \sum\limits_{j}^{d}
              \left(\partial_j \frac{\partial_j W}{2 W} \right) \right. \\
     {}  & + \left. \sum\limits_{k<l}^{d}
              \left( \{ -i \partial_k, \:
                   \frac{\partial_l W}{2 W} \}
             + 2 L_k \frac{\partial_l W}{2 W} \right) \tau_{kl}
           \right\},
\end{array}
\label{gh}
\end{eqnarray}
where $\tau_{kl} \equiv {i\over 2}[\tau_k,\, \tau_l]$, and
${\cal F}_{kl} \equiv \partial_k L_l - \partial_l L_k$. 
The vector potential $L_i$ naturally generates a gauge field
interaction structure in $d$-dimensional space so that ${\cal F}_{kl}$
may be seen as the strength of vector field.
The terms $ {\cal F}_{kl} \tau_{kl} $ and
$ \left\{ -i \partial_k, \: \frac{\partial_l W}{2 W} \right\} \tau_{kl}$
generalize the Pauli coupling and the orbit-spin coupling
interactions respectively. For the simple three-dimensional case,
these interpretations are more distinct. We take conveniently $\tau_j$ as
the Pauli matrices
\begin{eqnarray}
      \tau_1 = \left( \begin{array}{cc}
                         0  &  1   \cr
                         1  &  0
                      \end{array}
                \right ),
       \hspace{3mm}
      \tau_2 = \left( \begin{array}{cc}
                         0   &   i    \cr
                         -i  &   0
                      \end{array}
                \right ),
       \hspace{3mm}
      \tau_3 = \left( \begin{array}{cc}
                         1   &   0   \cr
                         0   &   -1
                      \end{array}
                \right ),
\end{eqnarray}
then the supersymmetric Hamiltonian (\ref{gh}) becomes
\begin{eqnarray}
\begin{array}{ll}
    H = & {1 \over 2} (\vec{p} + \vec{L})^2
           + {1 \over 2} k^2 W^2
           + {1 \over 2} \left( {\nabla W \over 2W} \right)^2
           + {1 \over 2} \nabla \cdot
                         \left( {\nabla W \over 2W} \right)
             \left( \begin{array}{cc}
                         -I_{4 \times 4}  &   0   \cr
                         0   &   I_{4 \times 4}
                      \end{array}
             \right )                         \\
    {}    & + {1 \over 2} \vec{B} \cdot \vec{\tau}
            - {1 \over 2} \left( {\nabla W \over 2W} \right)
              \times ( -i \nabla + \vec{L} ) \cdot \vec{\tau},
\end{array}
\label{h2}
\end{eqnarray}
where $\nabla$ is a three-dimensional gradient operator, 
$
  \vec{B} \cdot \vec{\tau} \equiv \nabla \times \vec{L} \cdot \vec{\tau}
$
and
$
  {\nabla W \over 2W} \times (-i \nabla )\cdot \vec{\tau}
$
are the usual Pauli coupling term and the orbit-spin coupling interaction
respectively. \cite{messiah} It can be seen that in three dimension
the new realized $N=4$ supersymmetric Hamiltonian (\ref{h2}) is an
eight-by-eight matrix, whereas the original one (\ref{h-2}) or 
(\ref{h-3}), in which $A_j$ and $D_j$ are taken as some appropriate 
first-order differential operator functions of the three-dimensional 
coordinates and momentum operators, is a four-by-four matrix.

Of course, in Eq. (\ref{ad-multi}), we may also take
$A_4= \sum_j (i p_j + \tilde{L}_j) \tilde{\tau}_j$, 
$D_4= \sum_j \tilde{F}_j \tilde{\tau}_j$, and $A_{\bar{n}}$, $\tilde{L}_j$
and $\tilde{F}_j$ are the functions of \{$x_n$\} as well. 
Thus, the Hermiticities of the supercharges $Q^{\pm}_{\alpha}$ require 
that $\tilde{L}_j$ and $\tilde{F}_j$ should be pure imaginary, and 
$\tilde{\tau}_i$ should be anti-Hermitian. 
The convenient choices $\tilde{L}_i = i L_j$, $\tilde{F}_j = i F_j$ and
$\tilde{\tau}_j = i \tau_j$ [i.e., $\tilde{\tau}_j$ constitute 
the Clifford algebra C(0,$d$) with negative metric] will lead to 
the same results as Eq. (\ref{gh}).

\section{APPLICATIONS}
\label{4}

In this section, using the diagonal matrix realization 
(\ref{diagonal}) of $N=4$ SSQM in one dimension, we shall study
in detail $N=4$ SCQM, which, discussed first 
by Fubini {\it et al.}, \cite{fr} is a generalization 
of $N=4$ SSQM, by considering two simple superpotentials 
$k/x$ and $\omega x$. 
Here the main task is to find a set of special exactly 
solvable potentials which can be brought into the framework 
of $N=4$ SCQM and the corresponding superalgebraic structures.

1. The first example is a one-dimensional superpotential
\begin{equation}
    W(x) = {k \over x},
\label{p1}
\end{equation}
where $k$ is a real constant, and $x\in (-\infty, \;\infty)$. 
Substituting Eq. (\ref{p1}) into Eq. (\ref{q-1d-matrix}) combined 
with $k^{\pm}=0$ and $k_3 = 1$ and Eq. (\ref{diagonal}), we may obtain 
respectively the supercharges 
\begin{eqnarray}
 \bar{Q}^{+}_{1} = \left( \begin{array}{cccc}
            0 & 0         & \bar{\eta}^{+}   &  0  \cr
            0 & 0         &        0   &  0  \cr
            0 & 0         &        0   &  0  \cr
            0 & \bar{\zeta}^{+} &        0   &  0  \cr
                    \end{array}
            \right), \hspace{4mm}
\bar{Q}^{+}_{2} = \left( \begin{array}{cccc}
            0          &  0 &       0  &  0  \cr
            0          &  0 & \bar{\eta}^{+} &  0  \cr
            0          &  0 &       0  &  0  \cr
            -\bar{\zeta}^{+} &  0 &       0  &  0  \cr
                    \end{array}
            \right),
\label{q-c-1}
\end{eqnarray}
where $ \bar{\eta}^{\pm} = p \pm i (k- { 1 \over 2 }) {1 \over x}$,
$ \bar{\zeta}^{\pm} = p \pm i (k+ {1 \over 2}) {1 \over x}$, and the
supersymmetric Hamiltonian
\begin{eqnarray}
\begin{array}{rl}
   H^{\mbox{sc}} & = {1 \over 2} p^2 + {1 \over 2 x^2}
        \left (
            \begin{array}{llll}
              k^2 - {1\over 4}   & {} & {} & {}  \\
              {} & k^2 - {1\over 4} & {} & {}  \\
              {} & {} & k^2 - 2k + {3\over 4} & {}   \\
              {} & {} & {} & k^2 + 2k + {3\over 4}
            \end{array}
         \right )   \cr
    {} & \equiv \mbox{diag} [H^{\mbox{c}}_1, \; H^{\mbox{c}}_2, \;
    H^{\mbox{c}}_3, \; H^{\mbox{c}}_4].
\end{array}
\label{sch-exm1}
\end{eqnarray}
It may be further verified that $H^{\mbox{sc}}$, together with
the dilatation generator $D$ and the conformal generator $K$,
which are given explicitly by
\begin{eqnarray}
D = - {1 \over 4} \{p,\,x \},  \hspace{4mm} K = {1 \over 2} x^2 \label{cdk}
\end{eqnarray}
fulfills the same commutation relations of the conformal algebra SO(2,1)
\cite{wybourne,br,hs}
\begin{eqnarray}
  [ D , \: H^{\mbox{sc}} ] = - i H^{\mbox{sc}},   \hspace{4mm}
  [ D , \: K ] =  i K,   \hspace{4mm}
  [ H^{\mbox{sc}} , \: K ] = 2 i D,
\label{ca}
\end{eqnarray}
as its four superpartner Hamiltonians $H^{\mbox{c}}_i$ ($i=1$, 2, 3, 4).
Hence, $H^{\mbox{sc}}$ given by Eq. (\ref{sch-exm1}) is the so-called
superconformal Hamiltonian, \cite{fr} which possesses not only the $N=4$ 
supersymmetry but also the dynamical conformal symmetry. 
Different from the results of Fubini {\it et al.}, 
\cite{fr} here we realize successfully a $N=4$ superconformal quantum
mechanics (SCQM) in one dimension. However, the realization of $N=4$ 
SCQM obtained in Ref. \cite{fr}, holding uniquely in two dimension, 
can not be reduced to the one-dimensional or extended to more than 
two-dimensional cases.
Furthermore, in the quartet structure of $H^{\mbox{sc}}$, three 
superpartner Hamiltonians $H^{\mbox{c}}_1(=H^{\mbox{c}}_2)$, 
$H^{\mbox{c}}_3$, $H^{\mbox{c}}_4$ are different, whereas in the quartet
structure in Ref. \cite{fr} only two different superpartner Hamiltonians, 
$H^{\mbox{c}}_1(=H^{\mbox{c}}_2)$ and $H^{\mbox{c}}_3(=H^{\mbox{c}}_4)$, 
appear.

Let the four-component spinor eigenfunction of $H^{\mbox{sc}}$ be
$\psi^{\mbox{sc}} = [\psi^{\mbox{c}}_{1},\: \psi^{\mbox{c}}_{2},\: 
\psi^{\mbox{c}}_{3}, \: \psi^{\mbox{c}}_{4}]^{\mbox{T}}$, where 
$\psi^{\mbox{c}}_{i}$ ($i=1$, 2, 3, 4) are respectively 
the eigenfunctions of $H^{\mbox{c}}_i$ belonging to the energy 
eigenvalues $E^{\mbox{c}}_i$. According to the transformation property
(\ref{trans-wfs}), $\psi^{\mbox{c}}_{i}$ are related by 
$\bar{\eta}^{\pm}$ and $\bar{\zeta}^{\pm}$ given in Eq. (\ref{q-c-1}).
In order to look for the eigenfunctions and energy eigenvalues
of $H^{\mbox{sc}}$, we check first whether or not a zero-energy ground
state exists by solving the following four first-order differential 
equations
\begin{eqnarray}
  \bar{Q}^{\pm}_{\mu} \, \psi^{\mbox{sc}}_{0} = 0,   
	\hspace{4mm} \mu =1, \, 2,
\label{ub}
\end{eqnarray}
where $\psi^{\mbox{sc}}_{0} \equiv 
[\psi^{\mbox{c}}_{0,1}, \; \psi^{\mbox{c}}_{0,2}, \; 
\psi^{\mbox{c}}_{0,3}, \; \psi^{\mbox{c}}_{0,4}]^{\mbox{T}}$ 
stands for a zero-energy eigenfunction. 
It is clear that neither of four solutions to Eq. (\ref{ub})
\begin{eqnarray}
\begin{array}{l}
   \psi^{\mbox{c}}_{0,1} \sim    x^{-(2k-1)/2},  \cr
   \psi^{\mbox{c}}_{0,2} \sim    x^{+(2k+1)/2},  \cr
   \psi^{\mbox{c}}_{0,3} \sim    x^{+(2k-1)/2},  \cr
   \psi^{\mbox{c}}_{0,4} \sim    x^{-(2k+1)/2}
\end{array}
\label{hg-function}
\end{eqnarray}
is normalizable on ($-\infty$, $\infty$) so that neither of
$H^{\mbox{c}}_i$ has zero-energy level. 
Hence, the supersymmetry of the $N=4$ superconformal quantum 
mechanical system described by $H^{\mbox{sc}}$ is broken.
From the SSQM point of view, we know that either of $H^{\mbox{c}}_i$
($i=1$, 2, 3, 4) has the same energy spectrum as the other three 
superpartner Hamiltonians, with $E^{\mbox{c}}_i$ being larger than 
zero. Consider the special case of $k=1/2$, it follows immediately 
from Eq. (\ref{sch-exm1}) that the conformal Hamiltonian 
$\breve{H}^{\mbox{c}}_4 = {1 \over 2}p^2 + {1\over x^2}$ is 
the superpartner of the Hamiltonian 
$\breve{H}_1={1 \over 2}p^2$ ($= \breve{H}_2 = \breve{H}_3$) 
of a free particle.
Therefore, the normalizable eigenfunctions of $H^{\mbox{c}}_i$ in
Eq. (\ref{sch-exm1}) corresponding to some positive definite energy
$E^{\mbox{c}}_i>0$ are the normalizable wave plane eigenfunctions, 
\cite{dff} i.e., the Bessel functions, $\psi^{\mbox{c}}_i = \sqrt{x}
J_{\lambda_i}(x \sqrt{2 E_i})$ ($i=1$, 2, 3, 4), with
$\lambda_i = k + (-)^i \left[ {i \over 3} \right]$.

Now let us establish a superalgebra that governs the above $N=4$ 
superconformal quantum mechanical system described by $H^{\mbox{sc}}$,
in which both SS(4) and SO(2,1) should be contained. 
Direct calculations show that the five generators of SS(4) and
three generators of SO(2,1) are not closed under the anticommutation
and commutation relations, for example, commuting the generators 
$\bar{Q}^{\pm}_{\mu}$ of SS(4) and the generator $K$ of SO(2,1) 
yields new operators
\begin{equation}
  S^{\pm}_{\mu}= x C^{\pm}_{\mu},  \hspace{4mm}
  \mu =1, \, 2.
\label{s}
\end{equation}
Thus, we obtain the following closed superalgebraic structure
\begin{eqnarray}
\begin{array}{l}
\begin{array}{llll}
1. &  \{ \bar{Q}^{+}_{\mu},\: \bar{Q}^{-}_{\nu} \}
           = 2 \delta_{\mu \nu} H^{\mbox{sc}},
   &  \{ \bar{Q}^{\pm}_{\mu},\: \bar{Q}^{\pm}_{\nu} \} =0,
   &  \mu, \: \nu = 1,\: 2;  				\cr
2. & \{ S^{+}_{\mu},\: S^{-}_{\nu} \}
          = 2 \delta_{\mu \nu} K,
   &  \{ S^{\pm}_{\mu},\: S^{\pm}_{\nu} \} = 0; 
   &  {} 							\cr
\end{array}
\\
\begin{array}{ll}
 3. & \{ \bar{Q}^{\pm}_{\mu},\: S^{\mp}_{\nu} \}
         = \delta_{\mu \nu} (k + (-)^{\mu} X_3 \pm 2 i D)
	+ 2 (\delta_{\mu \nu} -1 ) ( \delta_{\mu 1} X^{\pm} + 
		  \delta_{\mu 2} X^{\mp} ) ,         \cr
 {} & \{ \bar{Q}^{\pm}_{\mu},\: S^{\pm}_{\nu} \}  = 0;                  
\end{array}
\\
\begin{array}{llll}
 4. & [ D , \: H^{\mbox{sc}} ] = - i H^{\mbox{sc}},
    &  [ D , \: K ] =  i K,
    & [ H^{\mbox{sc}} , \: K ] = 2 i D;     
\end{array}
\\
\begin{array}{lll}
 5. &   [X_3 , \: X^{\pm}] = \pm 2 X^{\pm},
    &   [X^{+} , \: X^{-}] =  X_3;          
\end{array}
\\
\begin{array}{llll}
 6. & [H^{\mbox{sc}} ,\: \bar{Q}^{\pm}_{\mu}]=0,
    & [ D , \: \bar{Q}^{\pm}_{\mu}]
          = - {1 \over 2} i \bar{Q}^{\pm}_{\mu},
    & [ K , \: \bar{Q}^{\pm}_{\mu}  ]
          =  \pm 2 S^{\pm}_{\mu};        \cr
 {} & [ H^{\mbox{sc}} , \: S^{\pm}_{\mu} ]
          = \pm 2 \bar{Q}^{\pm}_{\mu},
    & [ D , \: S^{\pm}_{\mu}]
          =  {1 \over 2} i S^{\pm}_{\mu},
    &  [K ,\: S^{\pm}_{\mu}]=0;          \cr
 7. & [ H^{\mbox{sc}}, \: X_3 ] =0,
    & [ D, \: X_3 ] =0,
    & [ K, \: X_3 ] =0;                  \cr
 {} & [ H^{\mbox{sc}}, \: X^{\pm} ] =0,
    & [ D, \: X^{\pm} ] =0,
    & [ K, \: X^{\pm} ] =0;              \cr
\end{array}
\\
\begin{array}{lll}
 8. & [ X_{3} ,   \: \bar{Q}^{\pm}_{\mu} ] 
	= \pm (-)^{\mu + 1} \bar{Q}^{\pm}_{\mu},
    & [ X^{\pm} , \: \bar{Q}^{\pm}_{\mu} ] 
	= \pm (1 - \delta_{\mu 1}) \bar{Q}^{\pm}_{1},   \cr
 {} & [ X^{\pm} , \: \bar{Q}^{\mp}_{\mu} ] 
	= \mp (1 - \delta_{\mu 2}) \bar{Q}^{\mp}_{2};  
    & [ X_{3} ,   \: S^{\pm}_{\mu} ] 
	= \pm (-)^{\mu + 1} S^{\pm}_{\mu}                 \cr
 {} & [ X^{\pm} , \: S^{\pm}_{\mu} ] 
	= \pm (1 - \delta_{\mu 1}) \bar{S}^{\pm}_{1},
    & [ X^{\pm} , \: S^{\mp}_{\mu} ] 
	= \mp (1 - \delta_{\mu 2}) \bar{S}^{\mp}_{2}.
\end{array}
\end{array}
\label{sca}
\end{eqnarray}
We denote the above superalgebra by SC(4)$_1$, which is spanned by
eight fermionic generators $S^{\pm}_{\mu}$, $\bar{Q}^{\pm}_{\mu}$ 
($\mu = 1$, 2) and six bosonic generators $H^{\mbox{sc}}$, 
$D$, $K$, $X_3$ and $X^{\pm}$. 
The first equation in the second set of equations in Eq. (\ref{sca})
indicates that the fermionic generators $S^{\pm}_{\mu}$ may be seen
as square roots of the conformal generator $K$, which is similar as 
the supercharges are the square roots of the supersymmetric Hamiltonian. 
Similar to SO(2,1), the superconformal symmetry described 
by SC(4)$_1$ is dynamical since $H^{\mbox{sc}}$ does not
commute with $S^{\pm}_{\mu}$, $D$ and $K$. It is obvious that 
besides SS(4), SO(2,1), and SO(3)  
[see the fifth set of equations in Eq. (\ref{sca})], 
SC(4)$_1$ contains a Lie superalgebra OSp(2,1) as its 
subalgebra, \cite{kac} which, spanned by either
\{$S^{\pm}_1$, $\bar{Q}^{\pm}_1$, $H^{\mbox{sc}}$, $D$, $K$, $X_3$\} or 
\{$S^{\pm}_2$, $\bar{Q}^{\pm}_2$, $H^{\mbox{sc}}$, $D$, $K$, $X_3$\}, 
has been used to study $N=2$ SCQM. \cite{fr,dv}
Since the generators of SO(2,1) commute with 
those of SO(3) [see the seventh set of equations in 
Eq. (\ref{sca})], SC(4)$_1$ contains a maximum Lie 
subalgebra SO(3)$\times$SO(2,1). 
Consequently, we have the following two canonical group chains
\begin{eqnarray}
    \mbox{SC}(4)_1 \supset \left\{
\begin{array}{r}
    \mbox{OSp}(2,1) \supset \mbox{SO}(2) \times \mbox{SO}(2,1) \cr
    \mbox{SO}(3) \times \mbox{SO}(2,1)
\end{array}
    \right\} \supset
\begin{array}{ccc}
    {}           &   {}   &   {}          \cr
    \mbox{SO}(2) & \times & \mbox{SO}(2).  \cr
    {X_3}        &   {}   &  H^{\mbox{sc}}
\end{array}
\end{eqnarray}

2. The second example is a linear superpotential on the half line,
\begin{equation}
    W(x) = \omega x,
\label{p2}
\end{equation}
where $\omega$ is a real constant, and $x\in (0, \;\infty)$.
Substitution into Eq. (\ref{q-1d-matrix}) combined with $k^{\pm}=0$
and $k_3 = 1$ and Eq. (\ref{diagonal}) gives respectively 
the supercharges
\begin{eqnarray}
 \tilde{Q}^{+}_{1}
    = \left( \begin{array}{cccc}
            0 & 0         & \tilde{\eta}^{+}   &  0  \cr
            0 & 0         &        0   &  0  \cr
            0 & 0         &        0   &  0  \cr
            0 & \tilde{\zeta}^{+} &        0   &  0  \cr
             \end{array}
       \right), 
   \hspace{4mm}
\tilde{Q}^{+}_{2}
    = \left( \begin{array}{cccc}
            0          &  0 &       0  &  0  \cr
            0          &  0 & \tilde{\eta}^{+} &  0  \cr
            0          &  0 &       0  &  0  \cr
            -\tilde{\zeta}^{+} &  0 &       0  &  0  \cr
                    \end{array}
            \right) ,
\label{sq-2}
\end{eqnarray}
where $ \tilde{\eta}^{\pm} \equiv p \pm i (\omega x + { 1 \over 2 x }) $,
$ \tilde{\zeta}^{\pm} \equiv p \pm i (\omega x - { 1 \over 2 x })$, and
the corresponding supersymmetric Hamiltonian
\begin{eqnarray}
\begin{array}{rl}
    \tilde{H} & = {1 \over 2} p^2 + {1 \over 2} \omega^2 x^2  +
  \left (  \begin{array}{cccc}
      {3 \over 8 x^2}      & {}  & {}  & {}  \\
      {} & {3 \over 8 x^2} & {}  & {}  \\
      {} & {} & - {1 \over 8 x^2} + \omega & {}  \\
      {} & {} & {} & - {1 \over 8 x^2} - \omega
  \end{array}
  \right )             \\
     &  \equiv \mbox{diag}[\tilde{H}_1, \; \tilde{H}_2, \; 
		\tilde{H}_3, \; \tilde{H}_4].
\end{array}
\label{exm2-h}
\end{eqnarray}
Note that the set of potential functions $\tilde{V}_i$ 
($i=1$, 2, 3, 4) corresponding to the four superpartner 
Hamiltonians $\tilde{H}_i$ is different from the well-known 
radial harmonic oscillator potential 
$V_{\mbox{ho}}(l) = {1 \over 2} \omega^2 x^2 + { l(l+1) \over 2 x^2} $,
\cite{messiah} in which the angular momentum quantum number $l$ 
must be positive integer, though, from the point of view of mathematics,
$\tilde{V}_i$ in Eq. (\ref{exm2-h}) are the special cases of 
$V_{\mbox{ho}}(l)$ for $l$ taking some special values: 
$ \tilde{V}_1= \tilde{V}_2= V_{\mbox{ho}}(l= {1 \over 2} )$,
$ \tilde{V}_3= V_{\mbox{ho}}(l= - {1 \over 2}) + \omega$, and
$ \tilde{V}_4= V_{\mbox{ho}}(l= - {1 \over 2}) - \omega$.
It may be easily inferred by using Theorem X.7 in Ref. \cite{rs}
that all $\tilde{H}_i$ are hermitian on the half line (0, $\infty$)
since either of $\tilde{V}_i$ is in the limit point case in both 
zero and infinity.
Let the four-component spinor eigenfunction of $\tilde{H}$ be
$\tilde{\psi} = [\tilde{\psi}_{1},\: \tilde{\psi}_{2},\: 
\tilde{\psi}_{3}, \: \tilde{\psi}_{4}]^{\mbox{T}}$, where 
$\tilde{\psi}_{i}$ ($i=1$, 2, 3, 4) are respectively the eigenfunctions
of $\tilde{H}_i$ belonging to the energy eigenvalues $\tilde{E}_i$, and
are related by the four first-order differential operators 
$\tilde{\eta}^{\pm}$ and $\tilde{\zeta}^{\pm}$ given in Eq. (\ref{sq-2})
\begin{equation}
 \tilde{\psi}_{3} \stackrel{\tilde{\eta}^+}{\longrightarrow}
    \tilde{\psi}_{1} = \tilde{\psi}_{2}
    \stackrel{\tilde{\zeta}^+}{\longrightarrow}  \tilde{\psi}_{4}
 \hspace{3mm} \mbox{and} \hspace{3mm}
 \tilde{\psi}_{3} \stackrel{\tilde{\eta}^-}{\longleftarrow}
    \tilde{\psi}_{1} = \tilde{\psi}_{2}
    \stackrel{\tilde{\zeta}^-} {\longleftarrow}  \tilde{\psi}_{4}.
\end{equation}

We notice that the quantum mechanical system described by $\tilde{H}$ 
has no additional symmetry that can be, together with the $N=4$ 
supersymmetry, embedded in a larger supersymmetry
(for example the superconformal symmetry). 
If we rewrite $\tilde{H}$ as 
\begin{equation}
    \tilde{H} = \dot{H}^{\mbox{sc}}_0 + \omega^2 K + \omega Y_3,
\label{hh}
\end{equation}
where $\dot{H}^{\mbox{sc}}_0$ is the supersymmetric Hamiltonian 
given by Eq. (\ref{h-1d-matrix}) combined with $k_3 = k^{\pm} = 0$ 
and $W(x) = \omega x$, and $Y_3$ is a constant matrix, 
\begin{eqnarray}
    \dot{H}^{\mbox{sc}}_0 = {1 \over 2} p^2 +
  \left (  \begin{array}{cccc}
      {3 \over 8 x^2}      & {}  & {}  & {}  \\
      {} & {3 \over 8 x^2} & {}  & {}  \\
      {} & {} & - {1 \over 8 x^2} & {}  \\
      {} & {} & {} & - {1 \over 8 x^2}
  \end{array}
  \right ),
\hspace{4mm}
   Y_3 =  \left (
            \begin{array}{llll}
              0  & {} & {} & {}  \\
              {} & 0 & {} & {}  \\
              {} & {} & 1  & {}   \\
              {} & {} & {} & -1
            \end{array}
         \right ),
\label{exm2-h0y}
\end{eqnarray}
then using the same analysis employed in the last example, it is easy
to find that $\dot{H}^{\mbox{sc}}_0$ is a superconformal Hamiltonian 
since it satisfies not only SO(2,1), with the dilatation 
generator $D= -{1 \over 4} \{p,\,x \}$ and the conformal generator 
$K = {1 \over 2} x^2$, but also SS(4), with the four supercharges
$\dot{Q}^{\pm}_{\mu}$ ($\mu=1$, 2)
\begin{eqnarray}
 \dot{Q}^{+}_{1} = \left( \begin{array}{cccc}
            0 & 0         & p + {i \over 2x}  &  0  \cr
            0 & 0         &        0   &  0  \cr
            0 & 0         &        0   &  0  \cr
            0 & p - {i \over 2x}  &        0   &  0  \cr
                    \end{array}
            \right),
    \hspace{4mm}
 \dot{Q}^{+}_{2} = \left( \begin{array}{cccc}
            0          &  0 &       0  &  0  \cr
            0          &  0 & p + {i \over 2x}  &  0  \cr
            0          &  0 &       0  &  0  \cr
            -p + {i \over 2x}  &  0 &       0  &  0  \cr
                    \end{array}
            \right).
\end{eqnarray}
Introducing extra three operators
\begin{eqnarray}
   Y_3 = {1 \over 2} \sum_{\mu} (-1)^{\mu + 1}
            [C^{+}_{\mu}, \: C^{-}_{\mu}],
\hspace{4mm}
   Y^{\pm} = \pm C^{\mp}_{2} C^{\pm}_{1},        
\label{am-2}
\end{eqnarray}
which constitute SO(3) as well [see the fifth set of equations in 
Eq. (\ref{h0-scqma4})], 
it follows that $\dot{Q}^{\pm}_{\mu}$, $S^{\pm}_{\mu}$, 
$\dot{H}^{\mbox{sc}}_0$, $D$, $K$, $Y_3$ and $Y^{\pm}$ satisfy the 
following closed superalgebraic structure
\begin{eqnarray}
\begin{array}{l}
\begin{array}{llll}
1. &  \{ \dot{Q}^{+}_{\mu},\: \dot{Q}^{-}_{\nu} \}
           = 2 \delta_{\mu \nu} \dot{H}^{\mbox{sc}}_0,
   &  \{ \dot{Q}^{\pm}_{\mu},\: \dot{Q}^{\pm}_{\nu} \} =0,
   &  \mu, \: \nu = 1, \: 2;   			\cr
2. & \{ S^{+}_{\mu},\: S^{-}_{\nu} \}
          = 2 \delta_{\mu \nu} K,
   &  \{ S^{\pm}_{\mu},\: S^{\pm}_{\nu} \} = 0;          
   &  {}
\end{array}
\\
\begin{array}{lll}
3. & \{ \dot{Q}^{\pm}_{\mu},\: S^{\mp}_{\nu} \}
         =  \delta_{\mu \nu} (\pm 2 i D + Y_3),
    & \{ \dot{Q}^{\pm}_{\mu},\: S^{\pm}_{\nu} \}
      = (-)^{\mu} ( 1- \delta_{\mu \nu} ) 2 Y^{\mp};   
\end{array}
\\
\begin{array}{llll}
 4. & [ D , \: \dot{H}^{\mbox{sc}}_0 ] = - i \dot{H}^{\mbox{sc}}_0,
    & [ D , \: K ] =  i K,
    & [ \dot{H}^{\mbox{sc}}_0 , \: K ] = 2 i D;     
\end{array}
\\
\begin{array}{lll}
 5. &   [Y_3 , \: Y^{\pm}] = \pm 2 Y^{\pm},
    &   [Y^{+} , \: Y^{-}] =  Y_3;         
\end{array}
\\
\begin{array}{llll}
 6. & [\dot{H}^{\mbox{sc}}_0 ,\: \dot{Q}^{\pm}_{\mu}]=0;
    & [ D , \: \dot{Q}^{\pm}_{\mu}]
          = - {1 \over 2} i \dot{Q}^{\pm}_{\mu},
    & [ K , \: \dot{Q}^{\pm}_{\mu}  ]
          =  \pm  S^{\pm}_{\mu};        \cr
 {} & [ \dot{H}^{\mbox{sc}}_0 , \: S^{\pm}_{\mu} ]
          = \pm  \dot{Q}^{\pm}_{\mu},
    & [ D , \: S^{\pm}_{\mu}]
          =  {1 \over 2} i S^{\pm}_{\mu},
    &  [K ,\: S^{\pm}_{\mu}]=0;         \cr
 7. & [ \dot{H}^{\mbox{sc}}_0, \: Y_3 ] =0,
    & [ D, \: Y_3 ] =0,
    & [ K, \: Y_3 ] =0;                   \cr
 {} & [ \dot{H}^{\mbox{sc}}_0, \: Y^{\pm} ] =0,
    & [ D, \: Y^{\pm} ] =0,
    & [ K, \: Y^{\pm} ] =0;                \cr
 8. & [ Y_3 ,   \: \dot{Q}^{\pm}_{\mu} ] =
        \mp \dot{Q}^{\pm}_{\mu},
    & [ Y^{\pm} , \: \dot{Q}^{\pm}_{\mu} ] =
        \pm (-)^{\mu + 1} \dot{Q}^{\mp}_{\nu},
    & [ Y^{\pm} , \: \dot{Q}^{\mp}_{\mu} ] = 0;    \cr
 {} & [ Y_3 ,   \: S^{\pm}_{\mu} ] =
        \mp S^{\pm}_{\mu},
    & [ Y^{\pm} , \: S^{\pm}_{\mu} ] =
        \pm (-)^{\mu} S^{\mp}_{\nu},
    & [ Y^{\pm} , \: S^{\mp}_{\mu} ] = 0,
\end{array}
\end{array}
\label{h0-scqma4}
\end{eqnarray}
where $ S^{\pm}_{\mu}$ has been given by Eq. (\ref{s}). 
We denote the above superalgebra by SC(4)$_2$, which, 
different from SC(4)$_1$ defined by Eq. (\ref{sca}),
has the same subgroup structure as SC(4)$_1$, i.e.,
\begin{eqnarray}
    \mbox{SC}(4)_2 \supset \left\{
\begin{array}{r}
    \mbox{OSp}(2,1) \supset \mbox{SO}(2) \times \mbox{SO}(2,1) \cr
	                      \mbox{SO}(3) \times \mbox{SO}(2,1) 
\end{array}
    \right\} \supset
\begin{array}{ccc}
    {}           &   {}   &   {}          \cr
    \mbox{SO}(2) & \times & \mbox{SO}(2).  \cr
    Y_3          &   {}   &  \dot{H}^{\mbox{sc}}_0
\end{array}
\end{eqnarray}

To determine the eigenfunctions and energy eigenvalues of 
$\tilde{H}$ by algebraic method, it is convenient to regroup 
the previous operators $\dot{H}^{\mbox{sc}}_0$, $D$, $K$, 
$\dot{Q}^{\pm}_{\mu}$ and $S^{\pm}_{\mu}$ as
\begin{eqnarray}
\begin{array}{ll}
    T_3 = {1 \over 2 \omega} \dot{H}^{\mbox{sc}}_0 +
        { \omega \over 2 } K,
&   T^{\pm} = {1 \over 2 \omega} \dot{H}^{\mbox{sc}}_0
        - { \omega \over 2 } K \mp i D,           \cr
    L^{\pm}_{\mu}= {1 \over 2 \sqrt{\omega}}
        \dot{Q}^{\pm}_{\mu} -
        {\sqrt{\omega} \over 2 } S^{\pm}_{\mu},
&   R^{\pm}_{\mu}= {1 \over 2 \sqrt{\omega}}
        \dot{Q}^{\pm}_{\mu} +
        {\sqrt{\omega} \over 2 } S^{\pm}_{\mu}.
\end{array}
\label{sca4-2}
\end{eqnarray}
Note that $R^{\pm}_{\mu}$ ($\mu = 1$, 2) are up to
normalization constants the supercharges $\tilde{Q}^{\pm}_{\mu}$
associated with $\tilde{H}$, i.e., 
$R^{\pm}_{\mu} = {1 \over 2 \sqrt{\omega}} \tilde{Q}^{\pm}_{\mu}$. 
Owing to the fact $\tilde{H} = 2 \omega (T_3 + {1 \over 2} Y_3)$, 
there exists a simple relation between the energy eigenvalues 
$\tilde{E}_i$ ($i=1$, 2, 3, 4) of $\tilde{H}_i$ and the energy 
eigenvalues $e_i$ ($i=1$, 2, 3, 4) of $T_{i,3}$, the $i$th diagonal 
component of $T_3$, 
\begin{equation}
    \tilde{E}_i = e_i + \Delta_i, \hspace{4mm} 
    \Delta_i = (-)^{i+1} {1 \over 2} \left[ {i \over 3} \right] ,
\label{h-t3}
\end{equation}
and their corresponding eigenfunctions are identical.
Therefore, the eigenfunctions and energy eigenvalues of $\tilde{H}$
may be directly obtained provided that those of $T_3$ are known. 
With the help of Eq. (\ref{h0-scqma4}), the closed anticommutation 
and commutation relations satisfied by the six bosonic operators 
$T_3$, $T^{\pm}$, $Y_3$, $Y^{\pm}$ and eight fermionic operators 
$L^{\pm}_{\mu}$, $R^{\pm}_{\mu}$, read
\begin{eqnarray}
\begin{array}{l}
\begin{array}{lll}
1. & [ T_3 , \: T^{\pm} ] = \pm T^{\pm},
   & [ T^{+} , \: T^{-} ] =  - 2 T_3;                   \cr
2. &   [Y_3 , \: Y^{\pm}] = \pm 2 Y^{\pm},
   &   [Y^{+} , \: Y^{-}] =  Y_3;                        \cr
3. &  \{ L^{+}_{\mu},\: L^{-}_{\mu} \}
           = T_3 - {1 \over 2} Y_3,
  &  \{ L^{\pm}_{\mu},\: L^{\pm}_{\nu} \} =
    \{ L^{\pm}_{\mu},\: L^{\mp}_{\nu} \} = 0;  \cr
4. &  \{ R^{+}_{\mu},\: R^{-}_{\mu} \}
           = T_3 + {1 \over 2} Y_3,
   &  \{ R^{\pm}_{\mu},\: R^{\pm}_{\nu} \} =
    \{ R^{\pm}_{\mu},\: R^{\mp}_{\nu} \} = 0;  \cr
5. & \{ L^{\pm}_{\mu},\: R^{\mp}_{\nu} \}
         = \delta_{\mu \nu} T^{\mp},
  & \{ L^{\pm}_{\mu},\: R^{\pm}_{\nu} \}
         = (-1)^{\nu} (1- \delta_{\mu \nu}) Y^{\mp};   \cr
\end{array}
\\
\begin{array}{llll}
6. & [ T_3, \: Y^{\pm} ] =0,  & [ T_3, \: Y_{3} ] =0,
   & [ T^{\pm}, \: Y^{\pm} ] =0,                       \cr
{} & [ T^{\pm}, \: Y_{3} ] =0, & [ T^{\pm}, \: Y^{\mp} ] =0;
   & {}                                                \cr
7. & [ T_3 , \: L^{\pm}_{\mu} ]
    = \pm {1 \over 2} L^{\pm}_{\mu},
   & [ T^{\pm} , \: L^{\mp}_{\mu} ]
    = 0,
   & [ T^{\pm} , \: L^{\pm}_{\mu} ]
    = \pm L^{\mp}_{\mu};          \cr
{} & [ T_3 , \: R^{\pm}_{\mu} ]
    = \pm {1 \over 2} R^{\pm}_{\mu},
   & [ T^{\pm} , \: R^{\pm}_{\mu} ]
    = 0,
   & [ T^{\pm} , \: R^{\mp}_{\mu} ]
    = \mp L^{\mp}_{\mu};                \cr
8. & [ Y_3 , \: L^{\pm}_{\mu} ]
    = \mp L^{\pm}_{\mu},
   & [ Y^{\pm} , \: L^{\mp}_{\mu} ]
    = 0,
   & [ Y^{\pm} , \: L^{\pm}_{\mu} ]
    = \pm (-1)^{\mu +1} L^{\mp}_{\nu}; \cr
{} & [ Y_3 , \: R^{\pm}_{\mu} ]
    = \mp R^{\pm}_{\mu},
   & [ Y^{\pm} , \: R^{\mp}_{\mu} ]
    = 0,
   & [ Y^{\pm} , \: R^{\pm}_{\mu} ]
    = \pm (-1)^{\mu +1} L^{\mp}_{\nu},
\end{array}
\end{array}
\label{cr-2}
\end{eqnarray}
where $\mu, \: \nu = 1$, 2. 
We observe from Eq. (\ref{cr-2}) that (1) the first set of 
equations indicates that $T_3$, $T^{\pm}$ constitute SO(2,1)
also, where $T_3$ is a compact operator with a discrete spectrum, 
and $T^{+}$ ($T^{-}$) raises (lowers) the energy eigenvalues of 
$T_3$ by 1 unit.
(2) Similar to SC(4)$_1$, the superalgebra determined by 
Eq. (\ref{cr-2}) contains SO(2,1)$\times$SO(3) as its maximum 
Lie subalgebra as well, moreover, the values of $Y_3$
[see Eq. (\ref{exm2-h0y})] may be used to label the energy 
eigenvalues $e$ of $T_3$ or $\tilde{E}$ of $\tilde{H}$. 
(3) $T_3$ and $R^{\pm}_{\mu}$ (or $T_3$ and $L^{\pm}_{\mu}$,
$\mu=1$, 2) do not form SS(4), therefore, the energy spectrum
of $T_3$ is not four-fold degenerate.
(4) The first column of equations in the seventh and eighth sets of 
equations show that $R^{+}_{\mu}$ and $L^{+}_{\mu}$ raise 
the energy eigenvalues of $T_3$ by ${1 \over 2}$ unit meanwhile 
lower the values of $Y_3$ by $1$ unit, whereas $R^{-}_{\mu}$ and
$L^{-}_{\mu}$ lower the energy eigenvalues of $T_3$ by 
${1 \over 2}$ unit meanwhile raise the values of $Y_3$ by $1$ unit.

Now turn to the eigenfunctions and energy eigenvalues of $\tilde{H}$ 
by means of the similar approach as used in the last example.
Solving the following four equations
\begin{eqnarray}
  \tilde{Q}^{\pm}_{\mu} \, \tilde{\psi}_{0} = 0,  \hspace{4mm}
  \mu =1, \, 2,
\end{eqnarray}
where $\tilde{\psi}_{0} \equiv [\tilde{\psi}_{0,1}, \; 
\tilde{\psi}_{0,2}, \; \tilde{\psi}_{0,3}, \; 
\tilde{\psi}_{0,4}]^{\mbox{T}}$ stands for the zero-energy eigenfunction, 
gives rise to
\begin{eqnarray}
\begin{array}{c}
 \tilde{\psi}_{0,1} \sim {1 \over \sqrt{x}} \exp(- \omega x^2 / 2), \cr
 \tilde{\psi}_{0,2} \sim {1 \over \sqrt{x}} \exp(+ \omega x^2 / 2), \cr
 \tilde{\psi}_{0,3} \sim \sqrt{x} \exp(+ \omega x^2 / 2), \cr
 \tilde{\psi}_{0,4} \sim \sqrt{x} \exp(- \omega x^2 / 2).
\end{array}
\label{gs-2}
\end{eqnarray}
It is not difficult to find by simple calculations that of 
the four eigenfunctions $\tilde{\psi}_{0,i}$ ($i=1$, 2, 3, 4), 
only $\tilde{\psi}_{0,4}$ is square-integrable on 
the interval (0, $\infty$), whose normalized form is
\begin{eqnarray}
 \tilde{\psi}_{0,4} = \sqrt{\omega x } \exp(- \omega x^2 /2),
\label{4gs}
\end{eqnarray}
that is, only $\tilde{H}_4$ has a normalizable zero-energy ground state,
whereas the other three superpartner Hamiltonians 
$\tilde{H}_1$, $\tilde{H}_2$, $\tilde{H}_3$ do not have.
Of course, $\tilde{\psi}_{0,4}$ is also the normalizable ground state
eigenfunction of the fourth diagonal component $T_{4,3}$ of $T_3$, 
with its corresponding energy being larger than zero. 
With the help of the superalgebraic relations (\ref{cr-2}),
the eigenfunctions $\tilde{\psi}_{n',i}$ ($n'=1$, 2, ...; 
$i=1$, 2, 3, 4) for arbitrary excited states may be obtained from 
$\tilde{\psi}_{0,4}$ through two steps: 
first applying $n$ times the raising operator $T^+$ to 
$\tilde{\psi}_{0,4}$ produces all the excited states 
$\tilde{\psi}_{n',4}$ ($n'=1$, 2, ...) of $T_{4,3}$, and then acting 
respectively once and twice on $\tilde{\psi}_{n',4}$ with 
$\tilde{Q}^{-}_{\mu}$ or $R^{-}_{\mu}$ or $ L^{-}_{\mu}$ 
gives the other excited states $\tilde{\psi}_{n,i'}$ ($n=0$, 1, ...;
$i'=1$, 2, 3), i.e., all the eigenfunctions of $\tilde{H}_{i'}$ or
$T_{i',3}$. 
With the help of Rodrigus' formula for the generalized Laguerre 
polynomial $L^{a}_{n}(x)$ of positive integer $n$ and real parameter 
$a$ in argument $x$, we can finally obtain by induction 
\begin{eqnarray}
\begin{array}{l}
   \tilde{\psi}_{n,1} = 
	\sqrt{2 \omega x} \exp(- \omega x^2 /2) 
	L^{0}_{n}(\omega x^2)  ,   \cr
   \tilde{\psi}_{n,2} = 
	\sqrt{2 \omega x} \exp(- \omega x^2 /2) 
	L^{0}_{n}(\omega x^2)  ,   \cr
   \tilde{\psi}_{n,3} = 
	\sqrt{ { 2 x \over n+1}} (\omega x) \exp(- \omega x^2 /2) 
	L^{1}_{n}(\omega x^2)  ,   \cr
   \tilde{\psi}_{n,4} = 
	\sqrt{ { 2 x \over n+1}} (\omega x) \exp(- \omega x^2 /2) 
	L^{1}_{n}(\omega x^2)  ,   \cr
   n=0, \; 1, \; 2,... 
\end{array}
\label{ex2-wf}
\end{eqnarray}
Thus, the energy eigenvalues, $\tilde{E}_i$ and $e_i$ ($i=1$, 2, 3, 4)
related by Eq. (\ref{h-t3}), that correspond to the same normalized 
eigenfunctions (\ref{ex2-wf}), of $\tilde{H}_i$ and $T_{i,3}$ are
respectively
\begin{eqnarray}
\begin{array}{ll}
    \tilde{E}_1 = 2 \omega (e_1 + \Delta_1) = 2 \omega (n+1), 
	& \hspace{4mm} e_1 = n+1,  \cr
    \tilde{E}_2 = 2 \omega (e_2 + \Delta_2) = 2 \omega (n+1), 
	& \hspace{4mm} e_2 = n+1,  \cr
    \tilde{E}_3 = 2 \omega (e_3 + \Delta_3) = 2 \omega (n+1), 
	& \hspace{4mm} e_3 = n+1/2,  \cr
    \tilde{E}_4 = 2 \omega (e_4 + \Delta_4) = 2 \omega n,     
	& \hspace{4mm} e_4 = n+1/2,  \cr
	n=0, \; 1, \; 2,...
\end{array}
\label{el-2}
\end{eqnarray}
Eq. (\ref{el-2}) shows clearly that the $N=4$ supersymmetry of 
the quantum mechanical system described by $\tilde{H}$ is unbroken 
since its ground state energy is zero, and the four-fold degeneracies 
may be observed above the second level. However,
the quartet energy spectrum structure of $T_3$, which is not four-fold
degenerate, involves two sets of double degenerate spectra, moreover, 
the corresponding supersymmetry is broken since its ground state energy
is ${1 \over 2}$.
The energy spectrum structures of $\tilde{H}$ and $T_3$ are depicted
in Fig. 2 and Fig. 3, respectively. 
The solutions to the Schr\"odinger equations with the potentials
$a x^2 + {b \over x^2}$ for different $a$'s and $b$'s may be obtained 
by different approaches, see Refs. \cite{fr,wybourne,dff,fno,bs}.

\section{SUMMARY}

In this paper, we obtained the general form of $N=4$ SSQM
in arbitrary dimension, starting from the general form
of four supercharges, in which the fermionic degrees of freedoms 
include all the odd elements, $C_j$ and $C_j C_k C_l$, of 
the superalgebra associated with the Clifford algebra C(4,0).
Then, from them, we gave the one-dimensional physical realization
and the new multi-dimensional physical realization for the $N=4$ SSQM
by solving their respective constraint conditions. 
As applications, we studied in detail, on the base of the 
one-dimensional realization, two superconformal quantum 
mechanical systems with their superpotentials being
$k/x$ and $\omega x$, which possess both the $N=4$ supersymmetries 
and the dynamical conformal symmetries,
and established their corresponding superalgebraic 
structures, which are spanned by the eight fermionic generators and 
six bosonic generators. 
Our next work is to apply the general realizations of $N=4$ SSQM 
obtained in this paper to the other possible (quasi-) exactly 
solvable potentials, for example, listed in Refs. \cite{cks,ush}, 
and to discussing the $N=4$ supersymmetries in 
the relativistic quantum mechanical systems
such as Dirac equation, Klein-Gordon equation and so on.
From the point of view of mathematics, it is also 
of very interest to investigate the representations of SC(4)$_1$ and
SC(4)$_2$ and their relations to the classical Lie superalgebras
\cite{kac,bk}. This work is now under way.

\section*{ACKNOWLEDGMENTS}
This work is supported by Chinese National Natural Science Foundation
(19905005), Major State Basic Research Development Programs (G2000077400 
and G2000077604) and Tsinghua Natural Science Foundation.

\newpage

\begin{center}
{\bf \Large Captions}
\end{center}
Fig. 1. Typical structure of the four-fold degenerate energy 
spectrum of $\ddot{H}$ given by Eq. (\ref{diagonal}). 
The energy spectra of $H_{1}$, $H_{2}$, $H_{3}$, $H_{4}$ 
correspond to the $\beta =1$, $-1$, $0$, $0$ sectors, respectively. 
The eigenstate belonging to some energy level (dot) may be
connected with its left or right eigenstate  
by the supercharges $\ddot{Q}^{\pm}_{\mu}$ 
($\mu =1 $, 2), or, concretely, by the first-order
differential operators $\eta^{\pm}$ and
$\zeta^{\pm}$ along the horizontal line.
\vspace{5mm}  \\
Fig. 2.  Four-fold degenerate energy spectrum 
of $\tilde{H}$ given by Eq. (\ref{el-2}), which $N=4$
supersymmetry is unbroken since its ground state energy
is zero. Each eigenstate (dot) is connected with its surrounding 
eigenstates (dots) by the supercharges $\tilde{Q}^{\pm}_{\mu}$ 
($\mu=1$, 2) along the horizontal line, and by 
the raising/lowering operators $T^{\pm}$ along 
the vertical line. 
\vspace{5mm}  \\
Fig. 3. Quartet structure of the energy spectrum of $T_3$ 
related tightly to $\tilde{H}$ [see Eq. (\ref{el-2})]. 
Different from that of $\tilde{H}$ exhibited in Fig. 2,
$T_3$ possesses two sets of double degenerate spectra 
with the corresponding supersymmetry being broken.
Each eigenstate (dot) is connected with its surrounding 
eigenstates (dots) by the fermionic operators $R_{\mu}^{\pm}$, 
$L_{\mu}^{\pm}$ ($\mu = 1$, 2) along the slanting line,
and by the raising/lowering operators $T^{\pm}$ 
along the vertical line. 

\newpage

\unitlength=1mm
\begin{picture}(100,100)(10,10)
	\put(10,10){\vector(1,0){78}}
	\put(25,10){\line(0,1){1}}
	\put(21,6){$-1$}
	\put(75,10){\line(0,1){1}}
	\put(74,6){$1$}
	\put(49,6){$0$}
	\put(85,6){$\beta$}
	\put(50,10){\vector(0,1){78}}
	\put(52,84){$\ddot{E}$}

	\put(25,25){\circle*{2}}
	\put(25,40){\circle*{2}}
	\put(25,65){\circle*{2}}
	\put(25,25){\line(1,0){24}}
	\put(25,40){\line(1,0){24}}
	\put(25,65){\line(1,0){24}}
	\put(24,68){$\vdots$}
	\put(23,73){$\ddot{H}_2$}
	\put(35,40){\vector(1,0){2}}
	\put(35,44){$\zeta^+$}

	\put(48.8,15){\circle*{2}}
	\put(48.8,25){\circle*{2}}
	\put(48.8,40){\circle*{2}}
	\put(48.8,65){\circle*{2}}
	\put(48,68){$\vdots$}
	\put(44,73){$\ddot{H}_4$}

	\put(51.2,25){\circle*{2}}
	\put(51.2,40){\circle*{2}}
	\put(51.2,65){\circle*{2}}
	\put(51.2,25){\line(1,0){24}}
	\put(51.2,40){\line(1,0){24}}
	\put(51.2,65){\line(1,0){24}}
	\put(51.2,68){$\vdots$}
	\put(51.2,73){$\ddot{H}_3$}
	\put(61,40){\vector(1,0){2}}
	\put(61,44){$\eta^+$}

	\put(75,25){\circle*{2}}
	\put(75,40){\circle*{2}}
	\put(75,65){\circle*{2}}
	\put(75,68){$\vdots$}
	\put(73,73){$\ddot{H}_1$}


	\put(105,50){\circle*{2}}   
	\put(105,50){\vector(1,0){10}}  
	\put(105,50){\vector(-1,0){10}}

	\put(90,54){$\ddot{Q}^{-}_{1}$}
	\put(114,54){$\ddot{Q}^{+}_{1}$}
	\put(90,44){$ \ddot{Q}^{+}_{2}$}
	\put(114,44){$ \ddot{Q}^{-}_{2}$}

\end{picture}

\vspace{1cm}

\begin{minipage}{12cm}
Fig. 1. Typical structure of the four-fold degenerate energy 
spectrum of $\ddot{H}$ given by Eq. (\ref{diagonal}). 
The energy spectra of $H_{1}$, $H_{2}$, $H_{3}$, $H_{4}$ 
correspond to the $\beta =1$, $-1$, $0$, $0$ sectors, respectively. 
Each eigenstate (dot) may be connected with its left or right 
eigenstate by the supercharges $\ddot{Q}^{\pm}_{\mu}$ 
($\mu =1 $, 2), or, concretely, by the first-order
differential operators $\eta^{\pm}$ and
$\zeta^{\pm}$ along the horizontal line.
\end{minipage}

\newpage
\begin{picture}(100,100)(10,10)
	\put(10,10){\line(1,0){78}}
	\put(25,10){\line(0,1){1}}
	\put(21,6){$-1$}
	\put(75,10){\line(0,1){1}}
	\put(74,6){$1$}
	\put(49,6){$0$}
	\put(85,6){$Y_3$}
	\put(50,10){\vector(0,1){81}}
	\put(52,88){$\tilde{E}/2\omega$}

	\put(51.2,23){\vector(0,1){5.5}}
	\put(51.2,17){\vector(0,-1){6.5}}
	\put(50,19){1}
	\put(51.2,44){\vector(0,1){4.5}}
	\put(51.2,37){\vector(0,-1){6}}
	\put(50,39){1}
	\put(51.2,64){\vector(0,1){4.5}}
	\put(51.2,57){\vector(0,-1){6}}
	\put(50,59){1}
 
	\put(25,10){\circle*{2}}
	\put(25,30){\circle*{2}}
	\put(25,50){\circle*{2}}
	\put(25,70){\circle*{2}}
	\put(25,11){\line(0,1){19}}
	\put(25,30){\line(0,1){19}}
	\put(25,50){\line(0,1){19}}
	\put(25,30){\line(1,0){24}}
	\put(25,50){\line(1,0){24}}
	\put(25,70){\line(1,0){24}}
	\put(24,73){$\vdots$}
	\put(23,78){$\tilde{H}_4$}

	\put(48.8,30){\circle*{2}}
	\put(48.8,50){\circle*{2}}
	\put(48.8,70){\circle*{2}}
	\put(48.8,73){$\vdots$}
	\put(44,78){$\tilde{H}_2$}

	\put(51.2,30){\circle*{2}}
	\put(51.2,50){\circle*{2}}
	\put(51.2,70){\circle*{2}}
	\put(51.2,30){\line(1,0){24}}
	\put(51.2,50){\line(1,0){24}}
	\put(51.2,70){\line(1,0){24}}
	\put(51.2,73){$\vdots$}
	\put(51.2,78){$\tilde{H}_1$}

	\put(75,30){\circle*{2}}
	\put(75,50){\circle*{2}}
	\put(75,70){\circle*{2}}
	\put(75,30){\line(0,1){19}}
	\put(75,50){\line(0,1){19}}
	\put(75,73){$\vdots$}
	\put(72,78){$\tilde{H}_3$}


	\put(105,65){\circle*{2}}   
	\put(105,65){\vector(0,1){10}}  
	\put(105,65){\vector(0,-1){10}}
	\put(105,65){\vector(1,0){10}}  
	\put(105,65){\vector(-1,0){10}}

	\put(90,69){$ \tilde{Q}^{+}_{1}$}
	\put(114,69){$ \tilde{Q}^{-}_{1}$}
	\put(90,59){$ \tilde{Q}^{+}_{2}$}
	\put(114,59){$ \tilde{Q}^{-}_{2}$}
	\put(103,77){$ T^{+}$}
	\put(103,50){$ T^{-}$}

\end{picture}

\vspace{1cm}

\begin{minipage}{12cm}
Fig. 2.  Four-fold degenerate energy spectrum 
of $\tilde{H}$ given by Eq. (\ref{el-2}), which $N=4$
supersymmetry is unbroken since its ground state energy
is zero. Each eigenstate (dot) is connected with its surrounding 
eigenstates (dots) by the supercharges $\tilde{Q}^{\pm}_{\mu}$ 
($\mu=1$, 2) along the horizontal line, and by 
the raising/lowering operators $T^{\pm}$ along 
the vertical line. 
\end{minipage}

\newpage

\begin{picture}(100,100)(10,10)
	\put(10,10){\line(1,0){78}}
	\put(25,10){\line(0,1){1}}
	\put(21,6){$-1$}
	\put(75,10){\line(0,1){1}}
	\put(74,6){$1$}
	\put(50,6){$0$}
	\put(85,6){$Y_3$}
	\put(50,10){\vector(0,1){81}}
	\put(52,88){$e$}

	\put(50,20){\line(-1,0){2}}
	\put(48.8,13.5){\vector(0,-1){3}}
	\put(48.8,16.5){\vector(0,1){3}}
	\put(45,15){${1\over 2}$}

	\put(51.2,22){\vector(0,1){6}}
	\put(51.2,17){\vector(0,-1){6}}
	\put(50,19){$1$}
	\put(51.2,42){\vector(0,1){6}}
	\put(51.2,37){\vector(0,-1){6}}
	\put(50,39){$1$}
	\put(51.2,62){\vector(0,1){6}}
	\put(51.2,57){\vector(0,-1){6}}
	\put(50,59){$1$}

	\put(25,20){\circle*{2}}
	\put(25,40){\circle*{2}}
	\put(25,60){\circle*{2}}
	\put(25,20){\line(0,1){20}}
	\put(25,40){\line(0,1){20}}
	\put(25,40){\line(5,-2){23}}
	\put(25,60){\line(5,-2){23}}
	\put(25,80){\line(5,-2){23}}
	\put(24,63){$\vdots$}
	\put(23,70){$e_4$}

	\put(48.8,30){\circle*{2}}
	\put(48.8,50){\circle*{2}}
	\put(48.8,70){\circle*{2}}
	\put(48,73){$\vdots$}
	\put(43,79){$e_2$}

	\put(51.2,30){\circle*{2}}
	\put(51.2,50){\circle*{2}}
	\put(51.2,70){\circle*{2}}
	\put(51.2,30){\line(5,-2){23}}
	\put(51.2,50){\line(5,-2){23}}
	\put(51.2,70){\line(5,-2){23}}
	\put(51.2,73){$\vdots$}
	\put(51.2,79){$e_1$}

	\put(75,20){\circle*{2}}
	\put(75,40){\circle*{2}}
	\put(75,60){\circle*{2}}
	\put(75,20){\line(0,1){20}}
	\put(75,40){\line(0,1){20}}
	\put(74,63){$\vdots$}
	\put(72,70){$e_3$}

	\put(75,35){\line(0,1){20}}
	\put(74,60){$\vdots$}


	\put(105,65){\circle*{2}}   
	\put(105,65){\vector(0,1){10}}  
	\put(105,65){\vector(0,-1){10}}
	\put(105,65){\vector(2,-1){10}}
	\put(105,65){\vector(-2,1){10}}
	\put(90,73){$ R_{\mu}^{+}$}
	\put(90,63){$ L_{\mu}^{+}$}
	\put(115,63){$ R_{\mu}^{-}$}
	\put(115,54){$ L_{\mu}^{-}$}
	\put(103,77){$ T^{+}$}
	\put(103,50){$ T^{-}$}

\end{picture}

\vspace{1cm}

\begin{minipage}{12cm}
Fig. 3. Quartet structure of the energy spectrum of $T_3$ 
related tightly to $\tilde{H}$ [see Eq. (\ref{el-2})]. 
Different from that of $\tilde{H}$ exhibited in Fig. 2,
$T_3$ possesses two sets of double degenerate spectra 
with the corresponding supersymmetry being broken.
Each eigenstate (dot) is connected with its surrounding 
eigenstates (dots) by the fermionic operators $R_{\mu}^{\pm}$, 
$L_{\mu}^{\pm}$ ($\mu = 1$, 2) along the slanting line,
and by the raising/lowering operators $T^{\pm}$ 
along the vertical line. 
\end{minipage}


\newpage

\begin{thebibliography}{99}
\bibitem{nicolai} H. Nicolai, J. Phys. A {\bf 9}, 1497 (1976).
\bibitem{witten} E. Witten, Nucl. Phys. B {\bf 185}, 513 (1981).
\bibitem{kn} A. Kostelecky and M.M. Nieto, Phys. Rev. Lett. {\bf 53},
             2285 (1984); Phys. Rev. {\bf A 32}, 1293, 3243 (1985);
            Phys. Lett. {\bf 56}, 96 (1986).
\bibitem{zhang} J.Z. Zhang, Phys. Rev. Lett. {\bf 77}, 44 (1996).
\bibitem{rau} A.R.P. Rau, Phys. Rev. Lett. {\bf 56}, 95 (1986).
\bibitem{abcd} R.D. Amado, R. Bijker, F. Cannata, and J. Dedonder,
                 Phys. Rev. Lett. {\bf 67}, 2777 (1991).
\bibitem{shastry} B.S. Shastry, Phys. Rev. Lett.
                  {\bf 69}, 164 (1992).
\bibitem{ss} B.S. Shastry and B. Sutherland, Phys. Rev. Lett.
                  {\bf 70}, 4029 (1993).
\bibitem{messiah} A. Messiah, {\it Quantum Mechanics}
               (Wiley, New York, 1976). 
\bibitem{cks} F. Cooper, A. Khare, and U. Sukhatme, Phys. Rep.
                 {\bf 251}, 267 (1995).
\bibitem{junker} G. Junker, {\it Supersymmetric methods in quantum 
	and statistical physics} (Springer, Berlin 1996).
\bibitem{cr} M. De Crombrugghe and V. Rittenberg, Ann. Phys. (N.Y.)
             {\bf 151}, 99 (1983).
\bibitem{ry} V. Rittenberg and S. Yankielowicz, Ann. Phys. (N.Y.)
             {\bf 162}, 273 (1985).
\bibitem{jll} A. Jaffe, A. Lasniewski, and M. Lewenstein, Ann. Phys. (N.Y.)
             {\bf 178}, 313 (1987).
\bibitem{mathur} M. Mathur, Ann. Phys. (N.Y.) {\bf 204}, 233 (1990).
\bibitem{ft} A.M. Frydryszak and V.M. Tkachuk, quant-ph/0110049, 
	to appear in Czech. J. of Physics.
\bibitem{nepomechie} R. Nepomechie, Ann. Phys. (N.Y.) {\bf 158}, 67 (1984).
\bibitem{okubo} S. Okubo, J. Math. Phys. {\bf 32}, 1657 (1991).
\bibitem{kac} V.G. Kac, Adv. Math., {\bf 26}, 8 (1977).
\bibitem{sh} H.Z. Sun and Q.Z. Han, Progress in Physics, {\bf 3}, 81 (1983).
\bibitem{ih} L. Infeld and T.E. Hull, Rev. Mod. Phys. {\bf 23}, 21 (1951).
\bibitem{bd} J.D. Bjorken and S.D. Drell, {\it Relativistic Quantum
             Mechanics} (Mc Graw-Hill, New York, 1964).
\bibitem{abei} A. Andrianov, N. Borisov, M. Eides, and M. Ioffe, Phys. Lett. 
	A {\bf 109}, 143 (1985).
\bibitem{fr} S. Fubini and E. Rabinovici, Nucl. Phys. B {\bf 245}, 17 (1984).
\bibitem{wybourne} B.G. Wybourne, {\it Classical Groups for Physicists}
             (John Wiley \& Sons, New York, 1974).
\bibitem{br} A.D. Barut and R. Raczka, {\it Theory of Group Representations 
and Applications} (Polish Scientific Publishers, Warsaw, 1977).
\bibitem{hs} Q.Z. Han and H.Z. Sun, {\it Group Theory}
		(Peking University, Beijing, 1987).
\bibitem{dv} E. D'Hoker and L. Vinet, Commun. Math. Phys. {\bf 97}, 391 (1985).
\bibitem{rs} M. Reed and B. Simon, {\it Methods of Modern Mathematical Physics.
    II. Fourier Analysis, Self-Adjointness} (Academic Press, New York, 1975).
\bibitem{dff} V. de Alfaro, S. Fubini, and G. Furlan, 
              Nuovo Cim. A {\bf 34}, 569 (1976).
\bibitem{fno} D.J. Fern'andez, J. Negro, and M.A. del Olmo, Ann. Phys. (N.Y.)
     {\bf 252}, 386 (1996).
\bibitem{bs} V.G. Bagrov and B.F. Samsonov, Phys. Part. Nucl. 
	{\bf 28}, 374 (1997).
\bibitem{ush} A.G. Ushveridze, {\it Quasi-exact Solbable Models in Quantum 
	Mechanics} (Institute of Physics, Bristol, 1994).
\bibitem{bk} A. Bohm and M. Kmiecik, J. Math. Phys. {\bf 29}, 1163 (1988).
\end{thebibliography}
\end{document}